\documentclass[12pt,amsmath,amssymb]{iopart}

\usepackage{graphicx}

\usepackage{dcolumn}
\usepackage{bm}
\usepackage{units}
\usepackage{journames}

\usepackage[final]{LLNdef}
\usepackage[final]{modifications}

\usepackage{lineno}

\begin{document}

\renewcommand\topfraction{0.8}
\renewcommand\bottomfraction{0.7}
\renewcommand\floatpagefraction{0.7}



\topical[Growth modes of Fe(110) revisited\ldots]{Growth modes
  of Fe(110) revisited: a contribution of self-assembly to magnetic materials\\%
}

\author{O Fruchart, P O Jubert\footnote{Present address: IBM Almaden Research Center, 650 Harry Road, San Jose, CA-95120, USA}, M Eleoui,
  F Cheynis, B Borca, P David, V Santonacci, A Li{\'e}nard, M Hasegawa and C Meyer}%
\address{Laboratoire Louis N{\'e}el, CNRS, BP166, F-38042 Grenoble Cedex 9, France}
\ead{Olivier.Fruchart@grenoble.cnrs.fr}

\date{\today}

\begin{abstract}
We have revisited the epitaxial growth modes of Fe on W(110) and Mo(110), and propose an overview
or our contribution to the field. We show that the Stranski-Krastanov growth mode, recognized for
a long time in these systems, is in fact characterized by a bimodal distribution of islands for
growth temperature in the range $\sim\tempC{250-700}$. We observe firstly compact islands whose
shape is determined by Wulff-Kaischev's theorem, secondly thin and flat islands that display a
preferred height, \ie independant from nominal thickness and deposition procedure~($\thicknm{1.4}$
for Mo, and $\thicknm{5.5}$ for W on the average). We used this effect to fabricate self-organized
arrays of nanometers-thick stripes by step decoration. Self-assembled nano-ties are also obtained
for nucleation of the flat islands on Mo at fairly high temperature, \ie $\sim\tempC{800}$.
Finally, using interfacial layers and solid solutions we separate two effects on the preferred
height, first that of the interfacial energy, second that of the continuously-varying lattice
parameter of the growth surface.

\end{abstract}

\dataref{68.55.-a: Thin film structure and morphology; 81.16.Dn Self-Assembly; 81.15.Fg Laser
deposition; 75.75.+a Magnetic properties of nanostructures}

\submitto{\JPCM}

\maketitle

\vskip 0.5in

\vskip 0.5in


\section*{Introduction}

The first report of the epitaxial growth of the centered cubic element Fe along the (110)
orientation was published more than twenty years ago\cite{bib-GRA82}. Since that time a great
number of publications have followed on the subject, with so far no sign of vanishing interest or
slowing down of the discovery of new phenomena. The epitaxial growth under ultra-high vacuum~(UHV)
has mainly been performed on W and Mo, the two refractory metals than were shown to give rise to
no or marginal interface reactivity at any temperature up that of thermal
desorption\cite{bib-TIK90}, thus enabling one to fully explore the growth behavior. In this
article we will restrict ourselves to these two metals as a growth surface.

At first the interest arose for Fe(110)/W(110) because it appeared as a favorable system for an
experimental realization of a model 2D magnetic system. The reasons were that 1.~Fe and W are
immiscible in the bulk 2.~$(110)$ is the most dense plane of centered cubic materials, thus of
\apriori lowest surface energy and unfavorable against facet formation 3.~W(110) surface
energy~($\gamma_\mathrm{W}\sim\unit[4.0]{J.m^{-2}}$) is significantly larger than that of
Fe~($\gamma_\mathrm{Fe}\sim\unit[2.4]{J.m^{-2}}$), favoring wetting\cite{bib-VIT98}. The
immiscibility at the $(110)$ interface was confirmed for Fe/W at all temperatures\cite{bib-TIK90}.
The single atomic layer~(\AL) was found to be pseudomorphic, and was thus a model system for
calculations and analysis. The conditions are slightly less favorable for Mo, of surface energy
$\gamma_\mathrm{Mo}\sim\unit[3.4]{J.m^{-2}}$, and with alloys like $\mathrm{Fe}_2\mathrm{Mo}$ and
$\mathrm{Fe}_3\mathrm{Mo}_2$ reported in the bulk and for which indeed marginal reactivity was
reported at the Fe/Mo(110) interface above $\tempC{525}$\cite{bib-TIK90}. This explains why the
thorough investigation of epitaxial growth of Fe/Mo(110) started much later\cite{bib-MAL98},
finally revealing a behavior very similar to the case of Fe/W(110).

In short, Fe has a misfit close to \unit[10]{\%} with both Mo and W. For both Fe/Mo(110) and
Fe/W(110) it was reported that at any temperature the growth begins by a pseudomorphic first AL,
which above room temperature completes perfectly before the next layers start to grow. Then at
elevated temperature further growth proceeds in the Stranski-Krastanov mode, whereas continuous
although rough films are obtained at room temperature. More in detail, the following aspects have
been reported: essential interfacial inertness\cite{bib-TIK90,bib-KOL00}, atomic
diffusion~(W\cite{bib-REU98,bib-SLA04} and Mo\cite{bib-FRU03d}), the growth and structure of the
first \AL\cite{bib-GRA82,bib-GAR83,bib-ELM94,bib-BET95,bib-SAN99b,bib-MEY01}, the onset of
dislocation formation and the structure of the interfacial dislocation network as a function of
thickness~(W\cite{bib-GRA82,bib-BET95,bib-BOD96,bib-POP03} and
Mo\cite{bib-MAL98,bib-MUR02,bib-MAE04}), the growth of continuous films (including
characterization of stress and strain relaxation~(W\cite{bib-SAN99b,bib-MEY01,bib-POP03} and
Mo\cite{bib-CLE93,bib-FRU99c}), the kinetic roughness
 and a procedure for its smoothening\cite{bib-ALB93,bib-FRU98b,bib-KOH00},
 the formation of \unit[1]{\AL} and \unit[2]{\AL} stripes by
step-decoration of vicinal surfaces and as a function of miscut
azimuth~(W\cite{bib-ELM94,bib-HAU98,bib-HAU98b,bib-ELM00}), the onset of SK growth and
nanostructures obtained by annealing or directly by high temperature
growth~(W\cite{bib-BER90,bib-SAN98,bib-SAN98b,bib-SAN99,bib-WAC02,bib-ROH03,bib-USO05} and
Mo\cite{bib-MAL98,bib-MUR02}).

Some years ago we have considered Fe(110) magnetic systems in the context of the rising interest
in self-assembly, \ie the process by which nanostructures~(wires, dots \etc) can be spontaneously
fabricated by deposition on surfaces. In the following we will use the term
\emph{self-assembly}~(\SA) when the nanostructures display no order or only short-range order, and
\emph{self-organization}~(\SO) when the nanostructures display at least a medium range
order\cite{bib-FRU05b}. In search of optimized growth procedures for the fabrication of \SA or \SO
Fe(110) nanostructures we have undertaken a systematic investigation of growth processes in terms
of substrate material, amount of Fe deposited, growth temperature and possible annealing. In this
course we have gathered significant new data about the system, and uncovered new growth phenomena.
Notice that our evaporation technique is pulsed laser deposition performed under strict UHV
conditions. This technique yields results very similar to those obtained by molecular beam
epitaxy\cite{bib-SHE04,bib-FRU03d}. We indeed found no \emph{quantitative} discrepancy with
previously published data.

It is the purpose of this article to summarize our contribution to growth processes of Fe on
W(110) and Mo(110) surfaces. We have already published several articles fully or partly related to
the epitaxial growth of Fe(110). These concerned the principle of growth of high-quality buffer
layers on Sapphire\cite{bib-FRU98b}, the sub-monolayer range for Fe/Mo(110)\cite{bib-FRU03d}, the
occurrence of self-assembled Fe/Mo(110) compact islands and the deduction of the effective
interfacial energy for this system\cite{bib-FRU01b,bib-FRU01c}, the fabrication of
nanometers-thick stripes by step decoration for both Mo(110) and W(110)
surfaces\cite{bib-FRU02b,bib-FRU02c,bib-FRU04}. The present article aims at giving an overview of
our results, therefore some overlap with the above-cited articles is inevitable. Nevertheless we
report here more details and new results. Besides, we report preliminary results using a more
systematic approach. The first improvement consists in tuning continuously the in-plane lattice
parameter from that of W to that of Mo, using solid solutions. The second improvement consists in
controling independently from the previous point the chemical nature of the interface, either W or
Mo, making use of ultrathin pseudomorphic interfacial layers. Applied to the subsequent
fabrication of both islands or wires, this approach should in the future cast light on the
relevant parameters involved in the process of self-assembly of Fe(110). Incidentally, this latest
investigation has allowed us to further increase the versatility of Fe(110) nanostructures that
can be grown.

\section{Experimental procedures}
\label{sec-experimental}

The experimental growth setup has been significantly modified since its lastest
description\cite{bib-FRU01b,bib-FRU03d} so that we describe it here in detail. It consists of
three interconnected UHV chambers, each equipped with an ion pump with a Ti sublimator. The first
chamber is used for samples and targets storage, and samples preparation and analysis. It is
equipped with a  scanning Auger electron spectroscopy with MACII analyzer, scanning Argon ion
etching~(Thermo VG EX05), substrates outgassing (base pressure
$\unit[\scientific{2-3}{-10}]{Torr}$). A turbo pump is directly fitted to this chamber for use
during ion etching. The second chamber is equipped with a room-temperature Omicron STM-1 Scanning
Tunneling Microscope with a maximum field of view of $\unit[800]{nm}$~(base pressure
$\unit[\scientific{5}{-11}]{Torr}$). The growth proceeds in the third chamber~(base pressure
$\unit[\scientific{5}{-11}]{Torr}$). The evaporation method is pulsed-laser deposition~(PLD) using
a Quantel Nd-YAG frequency-doubled ($\lambda=\unit[532]{nm}$) laser with pulse width around
$\unit[10]{ns}$, a shooting frequency of $\unit[10]{Hz}$ and a maximum energy per shot of
$\unit[150]{mJ}$. The energy is adjusted by tuning the delay between the oscillator and the
amplification stage of the laser while the pumping stays constant. The growth chamber is equipped
with a $\unit[10]{kV}$ RHEED setup with a 10-bit CCD camera synchronized with laser shots, so that
RHEED patterns and oscillations can be monitored during deposition. The laser beam enters the
chamber through a window and impinges on the target at an angle \angledeg{27} away from the normal
to the surface. The main parameter for evaporation is the fluence $F$, \ie the energy per shot and
per unit area of the target\cite{bib-CHE93b,bib-TYR94,bib-SHE04}. $F$ can be adjusted with both
the laser power and a focusing lens located outside the chamber on the beam path. The focal length
of the lens is \unit[500]{mm}. The focus is located \emph{ahead} of the target, at a distance
$\unit[120-180]{mm}$ depending on the element to evaporate. The resulting working value lies in
the range $\unit[0.1-1]{J/cm^{2}}$, yielding a typical deposition rate of $\unit[0.5]{\AA/mn}$.
The laser is rastered on the target with the help of an \exsitu mirror mounted on an
electro-acoustic device. The substrate-to-target distance is $\unit[140]{mm}$. A Cu foil placed
before the sample can be moved continuously during deposition to mask part of the sample and thus
fabricate wedge-shaped films, \ie films whose thickness is varied from one place to another on a
sample. In the case of simple wedges the direction of the movement of the mask is always
progressively masking the sample. By rotating the sample around its axis by $\unit[180]{\deg}$ a
wedge with an opposite slope can be fabricated. To switch over to another target the beam can be
stopped with an \exsitu mechanical shutter, another target is moved in front of the beam, and the
focusing of the beam on the target is adjusted to a value suitable for each material. All these
movements are motorized and controlled through a computer that allows one to launch macros, like
the numerous opposite wedges necessary for the fabrication of solid solution buffer layers with
continuously varying in-plane lattice parameter, as reported in the next section. The location on
the wedge of the focused electron beam used for AES can be precisely and reproducibly controlled
owing to a translation stage. The convolution of the translation uncertainty and the size of the
spot on the sample is estimated to yield an uncertainty of $\thickmicron{100}$. For quantitative
AES the measurements are normalized to the beam intensity, which is measured regularly during each
series of measurements on a reference part of the sample consisting of either a thick deposit or a
bare buffer layer of a pure element.

All our film are deposited on commercial Sapphire \saphir\ wafers. Wafers from several suppliers
have been used~(Union Carbide, Bicron, Crystal GmbH), without noticing significant differences on
the crystal quality of the films grown on top. The miscut was found to vary from wafer to wafer,
however is unchanged over the two-inch area. Typical values lie below~\angledeg{0.1}. The sample
holder consists of a modified one inch Riber \emph{Molybloc} with a slot to insert an Omicron STM
holder. One of the clamps holding the substrate is rotated \insitu with the Omicron wobblestick to
scratch the surface of the sample and set an electrical contact for STM measurements. Openings
have been made in both the molybloc and the STM sample holder, so that the rear of the substrate
is in direct view of a Joule-heating filament. With this setup the sample size is
$\unit[6.5\times8.5]{mm}$, sitting on a $\thickmicron{500}$-wide ledge. To sit on this thin ledge
the two-inch Sapphire wafers are precisely laser-cut in a dedicated \exsitu chamber. To this
purpose the Nd-YAG laser is focused on the backside of the wafer, with a spherical lens
$f=\unit[200]{mm}$. A set of two cylindrical lenses with identical flat axes, one convergent and
one divergent however with shifted focal planes, is inserted before the focussing spherical lens.
This provides us with a beam with a very elongated elliptical shape, in turned transformed with a
diaphragm into a very narrow and nearly rectangular shape on the wafer, whose feature is a nearly
constant energy per unit \textsl{length}. With this design the wafer is cut much more efficiently
than with a simple spherical focusing, because laser ablation suffers from a saturating effect
above the evaporation threshold\cite{bib-CHE93b}. The linear cutting speed is then of the order of
one millimeter per second. The wafers are then cleaned in an ultrasonic bath in \unit[10-20]{\%}
diluted RBS25 (Chemical Products, Belgium), then rinsed with deionized water, to remove the Al
deposited at the back during the laser cutting. The rear of the wafer is sputter-coated by
sputtering prior to deposition with a \thicknm{100}-thick layer of a refractory metal, usually~W.
The purpose of this layer is to absorb most of the heating radiation emitted by the filament on
the deposition stage. Without this layer a significant amount of radiation is absorbed directly by
the epitaxial film, resulting in higher and uncontrollable temperatures. Substrate temperatures up
to \tempC{950} can be achieved with this setup, calibrated by an optical pyrometer and controlled
routinely  by a thermocouple in direct contact with the rear of the molybloc. Owing to this
indirect measurement of temperature, error bars on temperature reading might be significant,
especially in the range $\tempC{25-250}$ because of thermal inertia. The sample can also be
substituted with a quartz microbalance. Finally all deposits are capped before taking them back to
air. The capping, deposited at room temperature~(RT), may consist of $\thicknm{5}$ of Mo or W, or
of a few atomic layers of either covered by a \thicknm{2} of Al, Mg or Au. The latter two are
slightly annealed above room temperature to smoothen the surface. All these capping procedures
have been checked to result in a smooth layer displaying single atomic steps, and are an efficient
barrier against oxidation of Fe.

Atomic force microscope images were performed on a Park Scientific Instrument Autoprobe CP in
contact or non-contact mode. The Si tips are pyramidal (Mikromasch CSC21B, force constant
\unit[2]{N/m}). X-ray diffraction~(XRD) was performed using several setups equipped with a
monochromator. We worked essentially in a $\theta-2\theta$ geometry, probing either diffusion
vectors out of the plane, or in-plane with a grazing incidence. In the following misfits are
expressed with respect to the growing film, \eg
$\epsilon=(a_{\mathrm{Mo}}-a_{\mathrm{Fe}})/a_{\mathrm{Fe}}$. Finally, for element A grown on B we
use the notation A/B.

\section{Growth of buffer and interface layers}
\label{sec-buffer}

We do not use metal single-crystals as a substrate, but instead we grow bcc(110) buffer layers on
Sapphire \saphir$(11\overline20)$. On the one hand this is more time-demanding than using metal
single crystals as a new buffer layer has to be prepared for each new sample. On the other hand it
allows one to keep all samples and take them \exsitu for further characterization after growth.
Most importantly, this allowed us to vary the composition and thus tune the in-plane lattice
parameter, as will be shown in the following. We reported in the past the growth of high-quality
single-crystalline Mo(110)\cite{bib-FRU98b} and Nb(110) buffer layers\cite{bib-FRU02d}. After
recalling the growth procedure of such layers and give further characterization with respect to
Ref.\cite{bib-FRU98b}, we present three improvements that have been of use for the growth studies
of Fe(110) reported here, namely the successful suppression of twinning for W(110), the
fabrication of solid solutions with tunable in-plane lattice parameter, and the deposition of
atomically-flat pseudomorphic ultrathin layers over these buffer layers.

\subsection{Recall of the general procedure}
\label{subsec-procedureBuffers}

\begin{figure}
  \begin{center}
  \includegraphics[width=157mm]{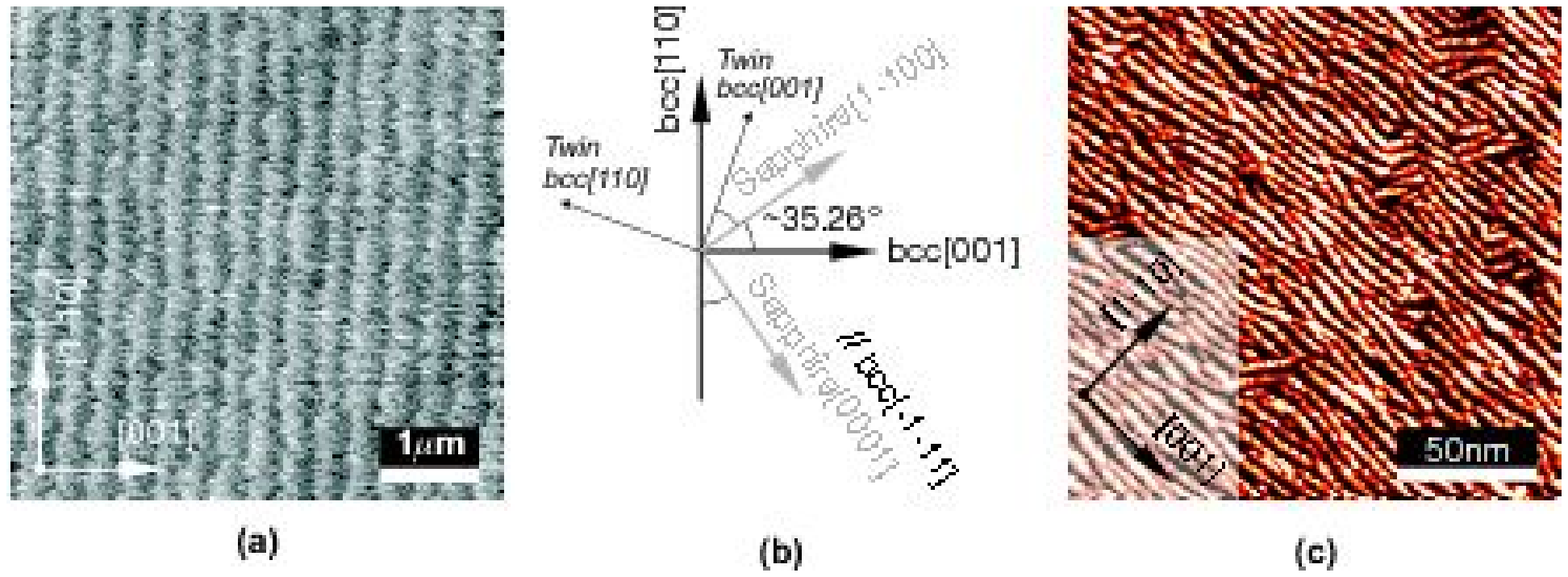}%
  \end{center}
  \caption{\label{fig-saphir-surface}(a)~\dataref{AFM/Cesar/FRU125/1013c02B.hdf. On BMP: scale 4A}$\unit[5\times5]{\mu\mathrm{m}}$
    AFM image of a typical
  $\saphir(11\overline20)$ surface displaying an array of monoatomic steps after
  annealing at $\tempC{800}$ for one hour under UHV. The vertical scale is \thickAA{4} (b)~epitaxial relationships between Sapphire
  and body centered cubic elements~(bcc), revealed by XRD and RHEED.
  The majority relationship\cite{bib-NIS86} is marked with
  dark lines and upright letters,
  while the minority twin is marked with thin lines and italicized letters
  (c)~The size of the crystallites of the minority twin are revealed in real space by STM
  by the areas with a $70^\circ$-rotated
  grooved pattern formed upon deposition at moderate temperature, see text.\dataref{bb53-m3.ori}}
\end{figure}

\begin{figure}
  \begin{center}
  \includegraphics[width=149mm]{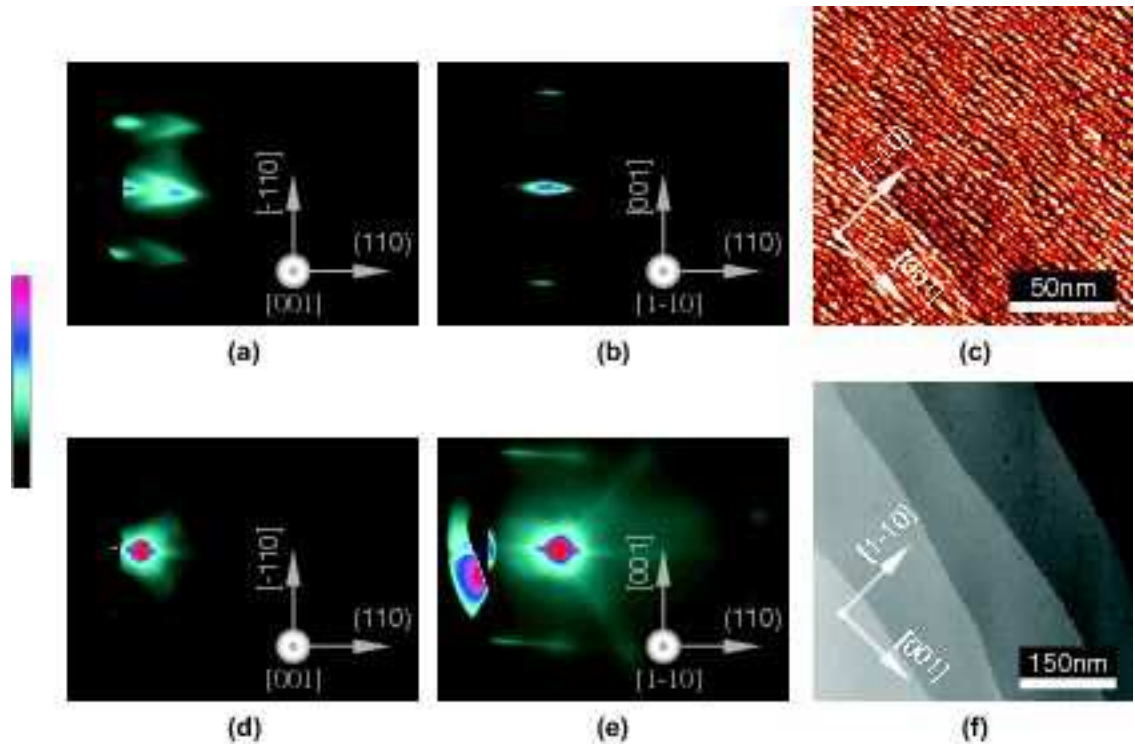}%
  \end{center}
  \caption{\label{fig-rough-one}Characterization of Mo(110)/\saphir$(11\overline20)$ films
    (a-c)~after deposition at room temperature and (d-f)~after deposition at
    room temperature followed by annealing at $\unit[800]{^\circ C}$.
  (a-b;d-e)~\dataref{FRU121}RHEED patterns along two azimuths (color scale on the left). The sample lies vertical at the left
  of the screen. (c,f)~STM images\dataref{BB54? and Fig.1.1 PhD POJ}.}
\end{figure}

The Sapphire wafers, prepared as reported above, are mounted on sample holders and outgassed under
UHV at $\tempC{800}$ for one hour and let cool down for one hour. At this stage the crystal
surface consists of atomically-flat terraces separated by single atomic steps, whose separation
and orientation are related to the residual miscut\bracketsubfigref{fig-saphir-surface}a. The
miscut was found to vary from wafer to wafer, however is unchanged over the two-inch area. A
typical value for the terrace width is \thicknm{200}~(see
FIG.\ref{fig-saphir-surface}a,\ref{fig-rough-one}f).

\begin{figure}
  \begin{center}
  \includegraphics[width=160mm]{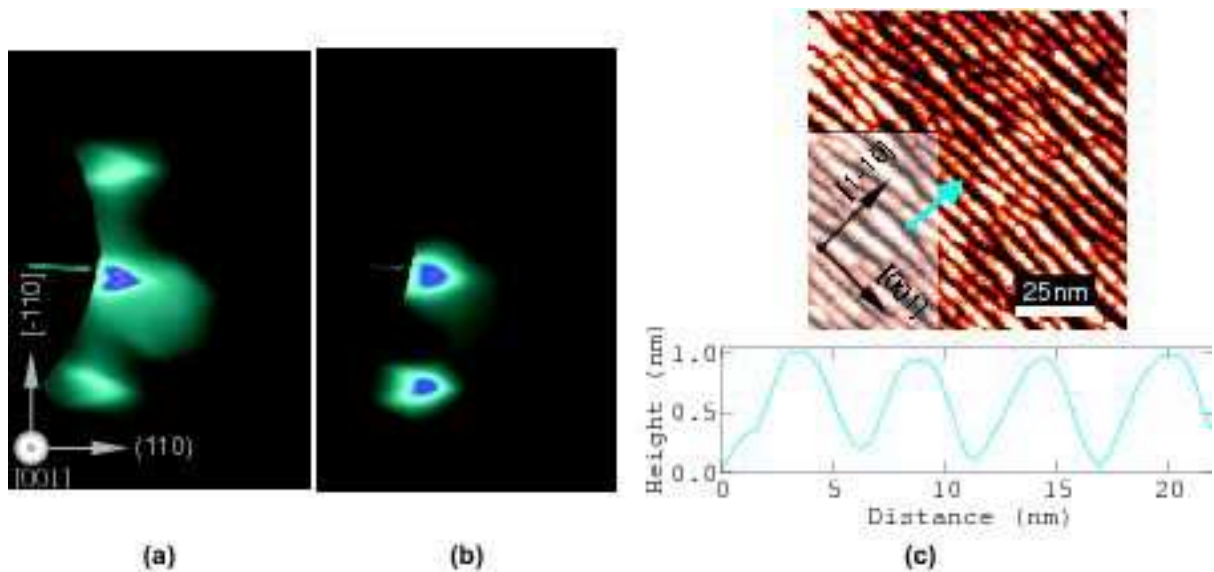}%
  \end{center}
  \caption{\label{fig-rough-two}\dataref{MH02 for RHEED, BB75m1 (apr{\`e}s multicouches Fe/W)}
  RHEED patterns on a grooved surface
  arising upon deposition of W(110) at \tempC{100}, along the (a)~[001] azimuth (b)~same as a,
  misoriented by \deg{1} away from [001] so as to enhance the satellite structure
  of the $[1-10]$ arrow-like streak (bottom). (c) STM
  image of the W(110) surface with a cross-section (arrow on the image).}
\end{figure}

Buffer layers consist of bcc materials deposited on Sapphire. The growth process described in the
following applies equally well for Mo, Nb and W, at least. The epitaxial relationship with
Sapphire agrees with that reported already long
ago\cite{bib-NIS86}\bracketsubfigref{fig-saphir-surface}b. These bcc materials are first deposited
at moderate temperature. In the past we were starting the growth at room temperature~(RT) and the
heater was switched on around $\thicknm{0.5-1}$ to reach $\tempC{150-200}$ by $\thicknm{1.5}$. We
nowadays perform the entire growth at RT, with no detectable difference in the quality of the
final annealed films, see next paragraph. Concerning the films before annealing, RHEED and STM
show that after growth at moderate temperature their surface is
rough\bracketsubfigref{fig-rough-one}{a-c}, displaying grooves oriented along $[001]$, a common
feature for bcc(110) epitaxy\cite{bib-ALB93,bib-FRU98b,bib-KOH00}. The microscopic origin of this
anisotropy of roughness has been discussed in Ref.\cite{bib-ALB93} and simulated in
\cite{bib-KOH00}, on the basis of anisotropic diffusion at the nucleation stage, followed by a
kinetic effect related to the Schwoebel barrier. In the former article the angle of the facets was
investigated for Fe(110) for a few thicknesses up to $\Theta=\thickAL{9}$, at \tempK{200}. It was
found that this angle increases during deposition to reach a value close to $\angledeg{18}$. The
authors could nicely reproduce their data with a model of layer-restricted-diffusion~(LRD) growth.
Our data partly confirm these conclusions. The arrow-like pattern of \figref{fig-rough-one}a is
blurred at the beginning of growth and sharpens upon deposition. However, whereas a LRD model
predicts a steady increase of the mean angle upon growth, we evidence that the angle of the facets
becomes stationary after some nanometers are deposited. RHEED and STM show in quantitative
agreement that the stationary angle is close to $\pm\angledeg{19}$, coherent with facets of type
$\{210\}$~(see Appendix~II). These values are consistent with the thickest deposits of Albrecht
\etal dealing with Fe\cite{bib-ALB93}, although these authors did not realize that a stationary
value had been reached. The breakdown of the continuous LRD model may be due to a modification of
the Schwoebel barrier, and/or to the modification of diffusion rates on the micro-terraces of the
facets of type $\{210\}$, due to finite size and/or the discreteness of the terrace. Indeed the
width of such a terrace is $\thickAA{\frac72a_{\mathrm{bcc}}\sqrt{2}/2\simeq7-8}$, and only three
atomic rows sit on each terrace. In fact in Ref.\cite{bib-KOH00} it was already suggested that
some type of facet should be stabilized, using geometrical arguments for impinging and diffusing
atoms. However it was claimed that the $\{310\}$ facets should be stabilized, contrary to our
findings. The stability of the $\{210\}$ facets has a positive effect on the array of grooves,
whose distribution of local orientation and period decrease upon further deposition after the
nominal facet angle has been reached on the average, \ie after $\thicknm{1-2}$. The array
progressively thus becomes self-organized, see \figref{fig-rough-two}.

The facetted grooves form upon deposition at temperatures below $\tempC{225}$\cite{bib-FRU98b} for
Mo and $\tempC{325}$ for~W. We have recently evidenced with RHEED that the roughness becomes
isotropic below approx. $\tempC{100}$ for W. A similar lower bound is likely to exist for Mo and
Fe, however shifted below RT in relation with the hierarchy of bonding strengths. In
Ref.\cite{bib-ALB93} facets were still observed for Fe at $\tempK{150}$. As our setup cannot
operate below RT we could not confirm this hypothesis. Nevertheless, the existence of a range of
temperatures for the kinetic formation of grooves, and its quantitative determination for W,
should be a valuable input for simulations, to determine the microscopic mechanism responsible for
this growth phenomenon. These grooves may for instance be a valuable template for the
self-organization of wires, like those obtained upon grazing-incidence ion
etching\cite{bib-RUS99,bib-VAL02,bib-SEK03}. With this in mind, let us add that STM investigations
revealed that the order seems to be at its best somewhere in the intermediate range of
temperatures, in terms both of alignement of the wires along [001], and of regular period along
$[1-10]$. The quality of the order is then reflected by the occurrence of a superstructure on the
RHEED patterns\bracketfigref{fig-rough-two}. The period deduced from RHEED is in accordance with
the period observed by STM, \ie around \thicknm{5} for Mo or W deposited at
$\tempC{100}$\bracketfigref{fig-rough-two}. The average depth of the grooves is $\thicknm1$ in
this case. In Ref.\cite{bib-ALB93} it was postulated that the period is selected during the
nucleation stage in the sub-atomic layer range of deposition, and accordingly was found to
increase when the temperature of deposition was increased. Although we have not investigated this
feature extensively, we have noticed that both nucleation and growth temperatures play a role.
Notice finally that all crossover temperatures between different growth modes may depend on the
deposition temperature, and concerning PLD depend also on the laser ablation parameters. For
instance, rather smooth films can be achieved at a temperature lower than the cross-over one for a
higher laser fluence, which is explained by the higher mobility of adspecies when they impinge on
the surface\cite{bib-FRU03d}. For an overview of differences between Molecular Beam Epitaxy~(MBE)
and PLD see Ref.\cite{bib-SHE04}.

As already reported\cite{bib-FRU98b}, there exists a dual epitaxial relationship of refractory
metals with \saphir$(11\overline20)$, both sharing $\mathrm{bcc}[111]\parallel\saphir[0001]$. The
dominant relationship is $\mathrm{bcc}[001]\sim\parallel\saphir[1\overline102]$\cite{bib-NIS86},
associated with intense streaks on RHEED, whereas weaker streaks on the RHEED pattern can be
attributed to a $\unit[70]{\deg}$ in-plane rotated relationship, in other words mirrored with
Sapphire $(1\overline100)$ plane. This was confirmed by grazing incidence XRD. The dual
relationship is explained by the initial completion of the dense atomic rows along bcc $<111>$
directions in the nucleation stage, followed by the completion of $<100>$ directions of atoms at
$\angledeg{\approx\pm35.26}$ on either side. The crystallites of the minority twin crystallites
are easily revealed upon growth at moderate temperature from the local direction of the grooved
pattern\bracketsubfigref{fig-saphir-surface}{c}. For Mo the minority crystallites cover an area
not exceeding $\unit[10]{\%}$ of the surface, with a typical lateral size of \thicknm{15}.

These layers are then annealed at $\tempC{800}$ for $\unit[30]{min}$ to flatten the surface and
improve the crystalline quality. RHEED patterns then display a sharp $1\times1$ pattern, and
reveal that the minority twin has vanished in the case of Mo, however not in the case of W~(upon a
similar annealing the minority twin vanishes in the case of Nb and V layers, however not in the
case of Ta. These three buffer layers have not be used for the overgrowth of Fe reported here). In
the case of Mo the disappearance of twins can also be checked by redeposition of a bcc material at
moderate temperature, in which case the array of grooves, locally aligned along [001], consists of
one single domain. It is this procedure that has been used for \figref{fig-rough-one}a-c. In
Ref.\cite{bib-SWI99} it was reported that Mo films deposited at $\tempC{800}$ do not show this
problem of twinning, however displayed a weak misorientation of crystallites in the as-deposited
state, resulting in micro grain boundaries. It was shown using Low-Energy Electron Microscopy that
these grain boundaries could be largely eliminated by annealing at $\tempC{1425}$ as they then
become very mobile. In our case we never came upon any evidence of the existence of such
micro-grain boundaries, in particular with STM. These annealed buffer layers consist of
atomically-flat terraces separated by mono-atomic surface steps\bracketsubfigref{fig-rough-one}f.
Large scale \exsitu AFM images show that these steps form a long-range ordered array on the
sample, whose orientation and period are linked with the residual miscut of the wafer after
polishing. Over a large number of wafers from various manufacturers the terrace size was found to
vary from $\thicknm{100}$ to $\thicknm{400}$, thus implying an average miscut angle around
$\angledeg{0.05}$. Notice that the miscut was always found to be uniform over the entire two-inch
wafer. Thus, as this wafer is cut into tens of sub-centimeter-size sub-wafers, once the miscut
angle and azimuth are known from microscopy, spare sub-wafers are stored that can be used whenever
a desired miscut is required.

\begin{figure}
  \begin{center}
  \includegraphics[width=78mm]{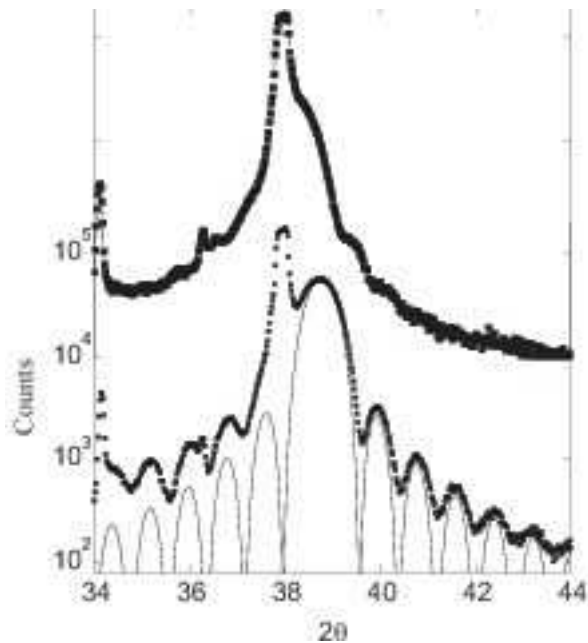}%
  \end{center}
  \caption{\label{fig-sapphire-kissing}XRD in the $\theta-2\theta$ geometry of a Nb film grown at $\tempC{200}$~(top)
    and annealed at $\tempC{800}$~(down). The dots stand for experimental points, while a qualitative fit of the Kissing fringes
    a shown in continuous line for the latter.\dataref{SP59-60}}
\end{figure}

The annealed buffer layers were characterized \exsitu with X-rays. The measurements with the
highest resolution were performed in the $\theta-2\theta$ geometry to measure the out-of-the-plane
$q$ vector. No detectable deviation from the bulk lattice parameter was evidenced, within the
accuracy of better than $0.1\%$. Kissing fringes with many orders are observed only after
annealing, resulting from the finite thickness of the layer, confirming the flatness of both
interfaces\bracketfigref{fig-sapphire-kissing}. For all films the crystalline coherence length
equals the film thickness minus a few atomic planes that contribute to an amorphous oxyde layer at
the free surface. In grazing incidence the epitaxial relationship was confirmed and the bulk
lattice parameter was retrieved, within the experimental accuracy of $0.5\%$.

\begin{figure}
  \begin{center}
  \includegraphics[width=140mm]{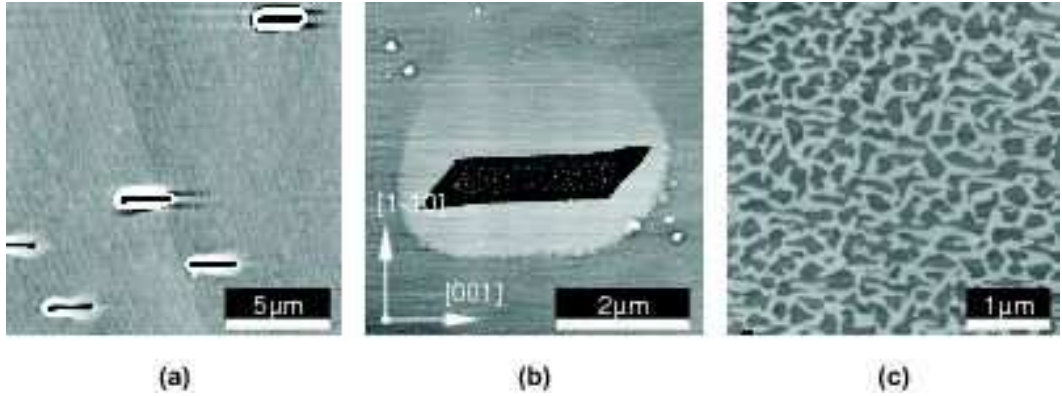}%
  \end{center}
  \caption{\label{fig-sapphire-unwet}Unwetting of Mo(\thicknm8)/$\saphir(11\overline20)$.
  (a)~After annealing at $\unit[900\pm50]{^\circ C}$. The array of monoatomic steps of Mo can be seen at the background.
  (b)~same conditions, with now the atomic steps running along [001]. The steps are not clearly seen
  because the scan was performed along the same direction. On both pictures
  the hole created by unwetting is decorated by a brim of Mo, with a flat surface, such that it
  appears as wedged over the slightly vicinal continuous Mo buffer layer. The small irregular patterns,
  looking like islands, are Al islands arising from a protective layer deposited under non-optimized conditions.
  (c)~A Mo film of nominal thickness $\Theta=\thicknm{2.2}$, which upon annealing at $\unit[900\pm50]{^\circ C}$
  unwets Saphir while forming nanostructures of height $\sim\thicknm{5.5}$.}
\end{figure}

The deposition and annealing parameters are not critical to get these buffer layers of high
cristalline and topographic quality. However a minimum nominal thickness $\Theta=\thicknm{8}$ is
required to avoid the unwetting of the substrate during annealing at $\tempC{800\pm50}$. The Mo
layers start to unwet sapphire for annealing temperatures above $\tempC{900\pm50}$ or
$\Theta<\thicknm{8}$\bracketsubfigref{fig-sapphire-unwet}a and complete unwetting occurs for a
significant overpassing of these parameters\bracketfigref{fig-sapphire-unwet}. We checked that
$\thicknm{50}$-thick buffer layers were stable up to at least $\tempdegC{1200}$, in agreement with
similar layers fabricated with MBE\cite{bib-SWI99}. Such buffer layers are now used by other
groups for the deposition of magnetic materials and nanostructures based on
Fe(110)\cite{bib-MAY01,bib-FON03,bib-VAL05,bib-FRA06}.

To conclude, the (110) layers deposited under the optimized procedure described above are of high
crystalline and topographic quality and are nearly indistinguishable from clean surfaces of metal
single crystals.

\subsection{Suppression of twinning for W}

It is important to eliminate the minority twins of W(110)/\saphir~to raise the quality of these
layers to the level of those of Mo, Nb and V. We indeed observed that Fe dots grown on a twinned
surface display features rotated by $\unit[70]{\deg}$ with respect to the expected $[001]$
direction, as a result of nucleation on the minority twin.

We could eliminate the minority twin by a so-called \emph{dusting} procedure: a small amount of Mo
is deposited first on Sapphire at room temperature, then W is deposited following the procedure
described above. RHEED reveals that after the growth of the dusting Mo plus W twins are present,
like in the case of pure elements deposited under identical conditions. However this time the
twinning vanishes upon annealing~(same procedure as above). The complete elimination of the twins
was confirmed by grazing-incidence X-rays and STM images of W overlayers deposited at moderate
temperature on annealed W(\thicknm{8})/Mo/\saphir buffer layers, similarly to the example shown on
\figref{fig-rough-one}a-c. It was found that as little as $\thicknm{0.6}$ is enough to suppress
the twinning. We did not attempt to determine the lowest required amount. The impact of ultrathin
dusting layers on the initiation of growth had been reported previously, \eg \thicknm{0.6} of V
deposited on MgO(001) are enough to change the growth of NiMnSb from polycrystalline to
epitaxial\cite{bib-TUR02b}. In our case the role of Mo in easing the recovery of a single crystal
through annealing has not been investigated. It might be to reduce the grain size of the minority
twin.

\subsection{Solid solutions}

\begin{figure}
  \begin{center}
  \includegraphics[width=83mm]{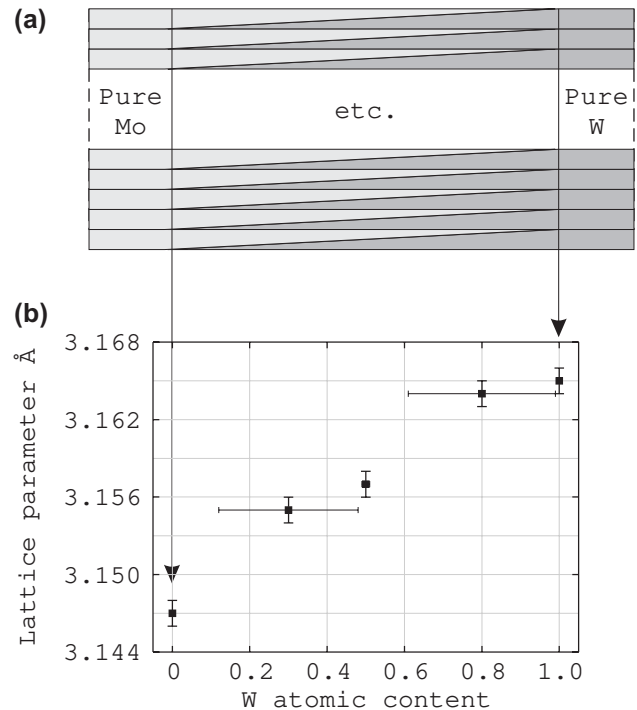}%
  \end{center}
  \caption{\label{fig-chemical-wedge}\dataref{Th{\`e}se Mustafa, Chapitre 02-ParametreDeMailleWxMo1-x.png, sans doute EM19}
    (a)~Schematic drawing of the geometry of a Chemical-Gradient Layer~(CGL)
    (b)~Out-of-plane lattice parameter of a Mo-W CGL. The vertical bars stand for experimental uncertainties,
    while the horizontal bars stand for the spread of composition on the slab measured related to the finite size of the area of CGL probed.}
\end{figure}

Lattice misfit is a key parameter in driving epitaxial growth modes. Thus it would be desirable to
continuously tune the misfit, which of course cannot be realized with pure elements. The
fabrication of bcc(001) buffer layers by codeposition of V and Nb was reported, which allowed the
authors to tune the in-plane lattice parameter of buffer layers in the full range between that of
V~($a=\thickAA{3.02}$) and that of Nb~($a=\thickAA{3.30}$)\cite{bib-FRI00}. We have developed a
process that allows us to fabricate a \emph{chemical-gradient layer}~(CGL), \ie a buffer layer
whose composition varies continuously from one end to the other end of the wafer. In the present
paper we focus on CGL made of mixtures of Mo and W, with bulk lattice parameters $\thickAA{3.147}$
and $\thickAA{3.165}$, respectively. The process is the following. Under the conditions detailed
in section \ref{subsec-procedureBuffers} we deposit sequentially W and Mo with the shape of
opposite wedges, with an area of pure element on each
side\bracketsubfigref{fig-chemical-wedge}{a}. The growth temperature protocol is identical to that
used for pure layers. The thickness of each bilayer is routinely $\thickAA{1}$, although periods
of $\thicknm{1}$ have been also used. The time required for the deposition of a full CGL is
doubled with respect to a pure-element layer, and reaches approximately four hours. Extra time is
indeed required to adjust the deposition parameters from one wedge to the next, consisting in
closing the beam stop, changing of target and of laser focusing on the target, rotating the sample
by $\angledeg{180}$ to alternate the direction of wedges, resetting the position of the mask, and
opening the beam stop. This procedure is fully automated and controlled through a computer.

Upon this deposition procedure performed at moderate temperature a dual epitaxial relationship is
evidence with RHEED, as with pure elements. Similarly to the case of W layers the minority twin
can be completely eliminated upon annealing, provided that a dusting layer of Mo is deposited
prior to the multilayer. After annealing the RHEED then displays a sharp $1\times1$ pattern. STM
reveals a mono-atomic-stepped surface, similar to that of pure
elements\bracketsubfigref{fig-rough-one}f. $\theta-2\theta$ XRD was performed along the
out-of-plane direction on several narrow slabs of these CGL. The out-of-plane lattice parameter
varies indeed from that of W to that of Mo. It is expected that the in-plane lattice parameter
undergoes also a linear variation.

Although these CGL are fabricated by a multilayer process, a fine intermixing is expected owing to
several arguments: the sub-monolayer period, the moderate deposition temperature, the small
lattice misfit and the total solubility of W-Mo. Accordingly we refer to these buffer layers as
\emph{solid solutions} throughout the manuscript. Preliminary results suggest that the process
works also finely for CGL made of Nb and Mo, two elements that also mix as solutions and form no
alloys with specific stoichiometry in the bulk. In this case RHEED suggests a flat surface of high
quality, with an in-plane lattice parameter continuously varying across the CGL. The process
should be extended in the near future to (V,Mo) CGLs, with a view to spanning the in-plane lattice
parameter of buffer solutions in the entire range $[\thickAA{3.02-3.30}]$.

\subsection{Pseudomorphic interfacial layer}

\begin{figure}
  \begin{center}
  \includegraphics[width=109mm]{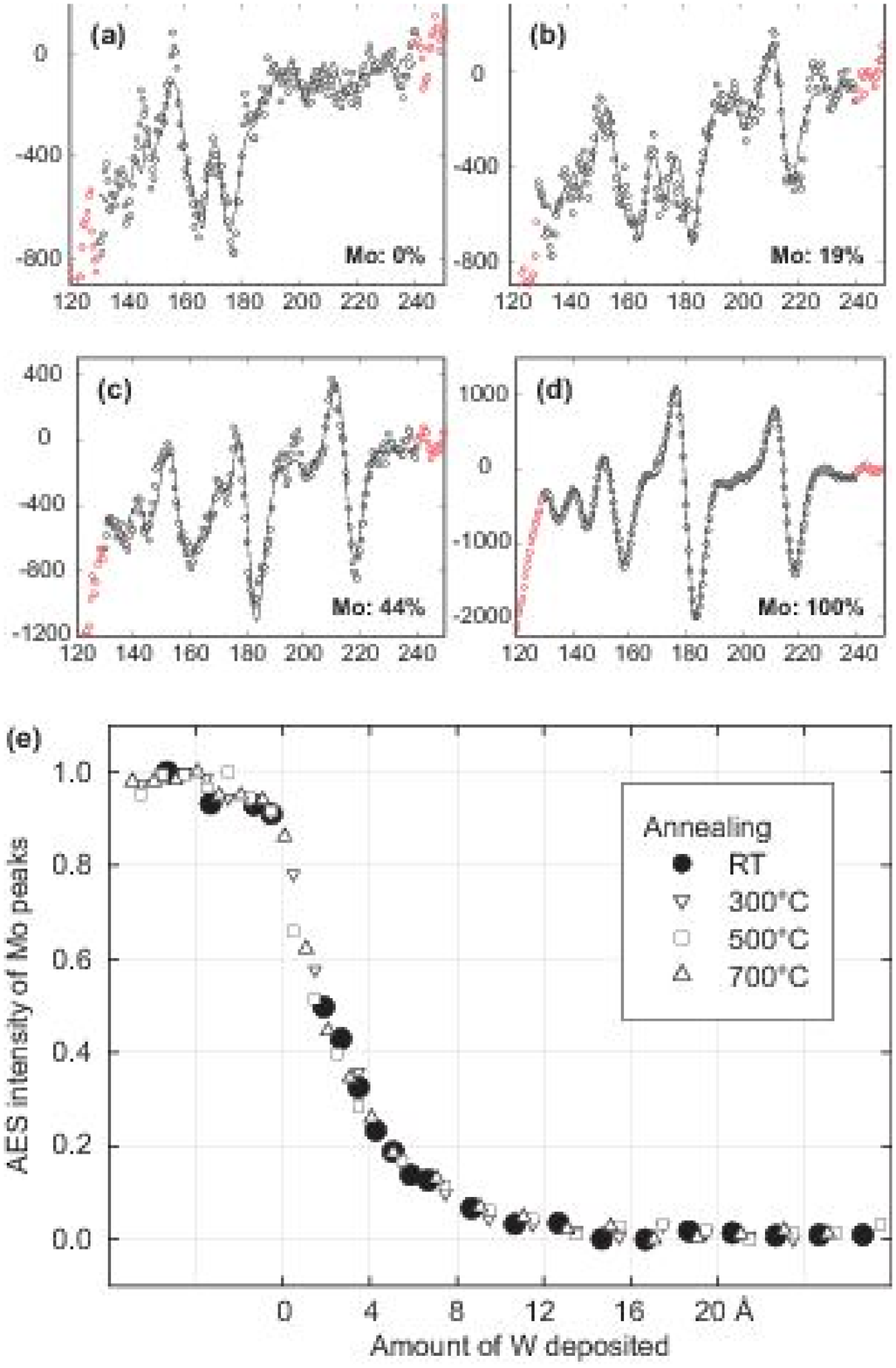}%
  \end{center}
  \caption{\label{fig-interfacial-layer}\dataref{EM14 for 19, 16.50,16,14mm}(a-d)~Selected AES spectra
    performed at RT in the range $[120-250\mathrm{eV}]$, of a wedged deposit of W on a smooth Mo(110) buffer
    layer~(normalized arbitrary units). Experimental data~(dots) are fitted in the range $[130-240\mathrm{eV}]$ with a linear combination
    of Mo and W reference spectra~(line). The fitted contribution of the Mo reference spectrum
    is shown in inset (e)~Intensity of the Mo spectrum as a function of the coverage
    of W, up to above $\thicknm{2.5}$, and following various annealing}
\end{figure}

The chemical nature of the interface is another key parameter in determining growth modes in
epitaxy. Ideally, it is desirable to control the chemical nature of the supporting surface
independently from the in-plane lattice parameter. To achieve this we have optimized a growth
process to fabricate an ultrathin pseudomorphic layer of either Mo or W on the buffer layers
previously described. We illustrate this procedure in this section for W deposited on
Mo(\thicknm{8}). A wedge of W is deposited at RT on a flat Mo buffer layer, followed by annealing
at increasing temperatures up to $\tempC{700}$. Auger spectra of the low energy peaks of W and Mo
were taken at RT after growth and after each step of
annealing\bracketsubfigref{fig-interfacial-layer}{a-d}. A quantitative analysis of the spectra in
the range $\unit[130-240]{eV}$ suggested that upon annealing no change occurs in the relative
intensity of both elements upon the entire range of
thickness\bracketsubfigref{fig-interfacial-layer}{e}. It is reasonable to assume that no
intermixing occurs upon RT deposition, first because of the high temperature of melting of both
elements, second because they have a similar atomic radius making exchange processes not clearly
favorable, and third because (110) is a dense plane of the centered-cubic structure. Thus we
assume that the spectra after deposition are a reference for a sharp interface. This is confirmed
by the nearly exponential decay of the Mo intensity as a function of W thickness. Auger spectra do
not show significant changes upon annealing, which suggests that the interface with Mo remains
sharp, and that no diffusion of Mo to the free surface occurs. This point is essential to control
the chemical nature of the interface. Finally we checked by STM that the surface resulting from
annealing was flat at the atomic scale. For subsequent growth of Fe we used layers of thickness in
the range $\thickAL{3-5}$, where it is expected that the layers remain pseudomorphic thanks to the
lattice misfit between Mo and W being smaller than $1\%$. This point could no be checked directly
due to the limited resolution of RHEED.

\section{Overview of growth processes of Fe/bcc(110)}

As recalled in the introduction, a large number of reports have already been made concerning the
UHV growth modes of epitaxial Fe(110), mainly on W and more recently on Mo. Both cases were
reported to be very similar, being of Stranski-Krastanov~(SK) type with one single AL of wetting.
Continuous films were successfully fabricated by RT deposition followed by moderate annealing, or
by rising the temperature during deposition. Nanostructures were observed to form upon annealing
at higher temperature or direct deposition at high temperatures.

\begin{figure}
  \begin{center}
  \includegraphics[width=157mm]{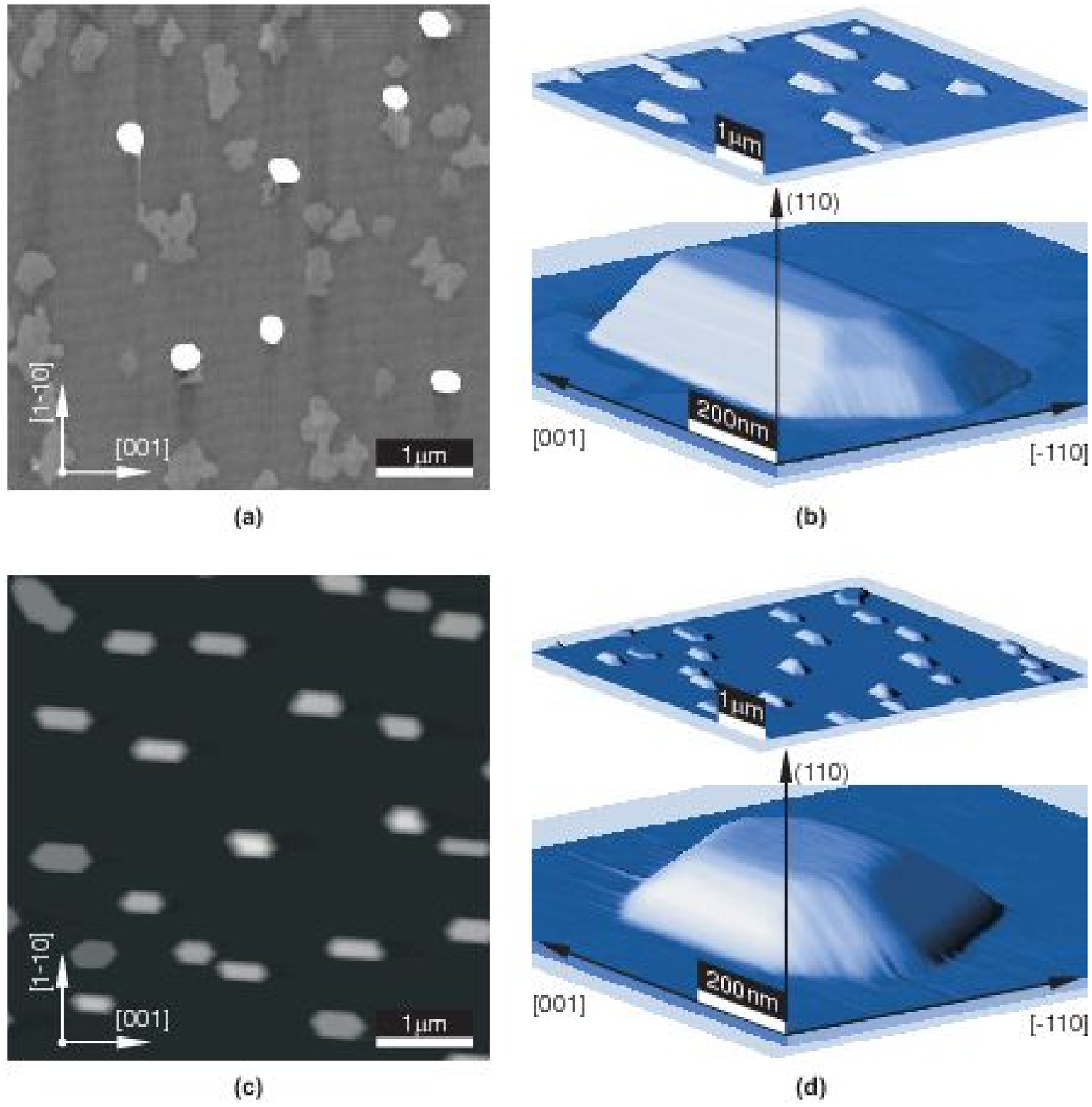}%
  \end{center}
  \caption{\label{fig-bimodal-growth}\dataref{(a)~Fig2.13 th{\`e}se POJ, Fru102 (11290040.hdf)
      (b)~Fru136b, 0404a03e.hdf et 0404a040.hdf (c)~FRU137b 0404a009.hdf (d)~FRU137b
      0404a009.hdf et 0404a02c.hdf}Illustration of the growth of Fe on (a-b)~Mo(110) and (c-d)~W(110).
      (a)~Deposition of $\Theta=\thicknm{0.4}$ (early stages of growth) on Mo(110)
      at $\unit[450]{^\circ\mathrm{C}}$ The grey scale is chosen so as to highlight the bimodal growth mode
      (b)~3D view and closeup view of Fe/Mo(110) compact
      dots, deposited
      on Mo(110) at $\unit[500]{^\circ\mathrm{C}}$, nominal thickness $\thicknm{2.5}$ (1:1 vertical scale)
      (c)~Deposition of $\Theta=\thicknm{2.5}$ on W(110) at $\unit[450]{^\circ\mathrm{C}}$
      (d)~3D view and closeup view of Fe/W(110) compact dots as in (c).}
\end{figure}

\begin{figure}
  \begin{center}
  \includegraphics[width=157mm]{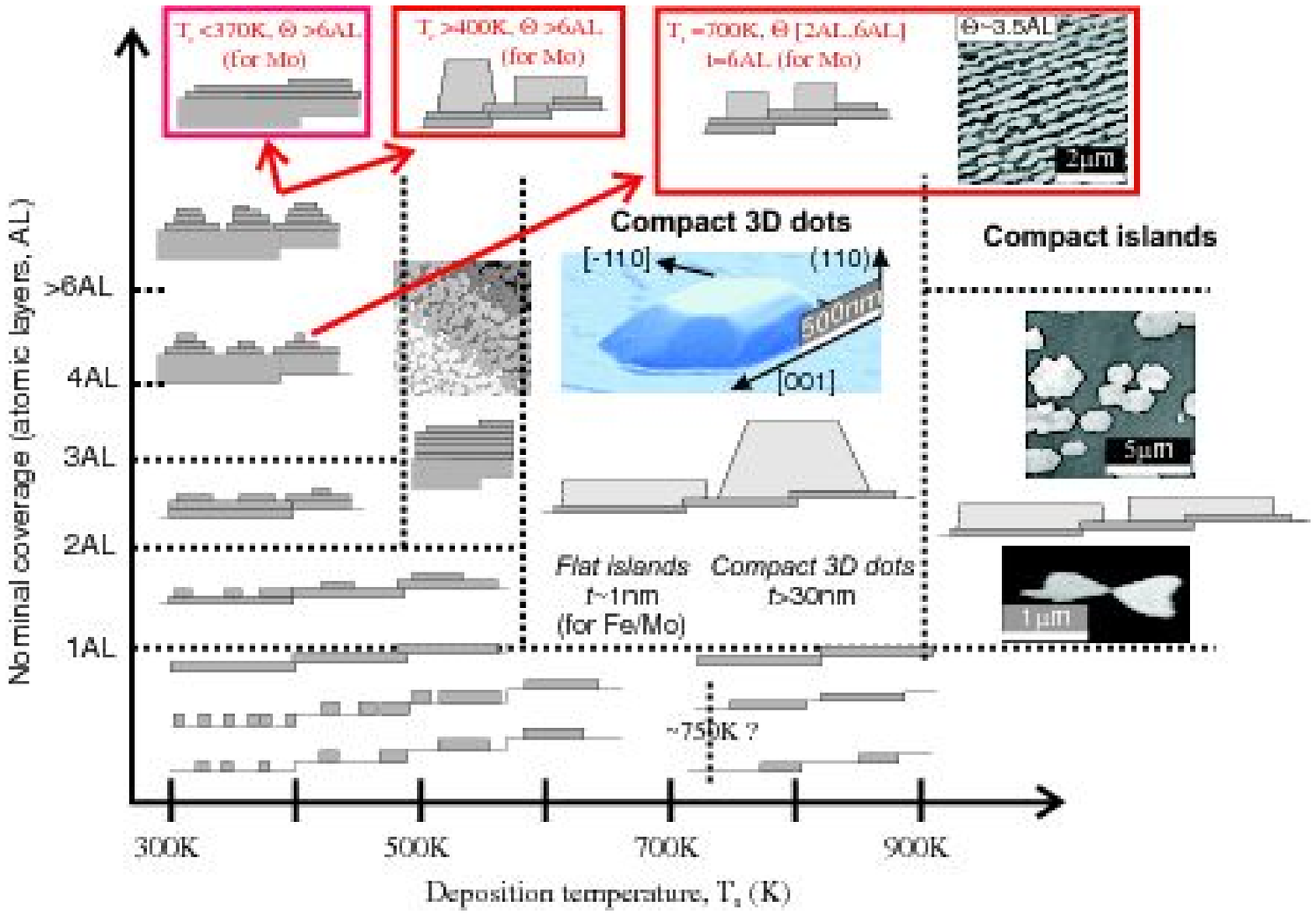}%
  \end{center}
  \caption{\label{fig-growth-overview}\dataref{Poster JSI2006}Schematic overview of growth modes
  of Fe on Mo(110) and W(110) surfaces, as a function of temperature of deposition and amount of
  Fe deposited. The output of annealing is displayed in rectangles. The growth modes
  for which we reported the first contribution are illustrated with a picture}
\end{figure}

We have uncovered that the SK mode is in fact characterized by a bimodal distribution of
islands\bracketfigref{fig-bimodal-growth}. The first type consists of compact islands~(\ie islands
with all three dimensions of the same order of magnitude) that grow nearly homothetically during
deposition, and whose shape can be understood on the basis of the Wulff-Kaischev geometrical
construction~(see \cite{bib-MUL00} and included references). The second type consists of flat and
thin islands (the ratio of lateral over vertical size can exceed 100), which display a nearly
mono-disperse height that depends very little on deposition conditions nor on the lateral size of
the islands. We already reported this bimodal growth mode for Fe/Mo(110)\cite{bib-FRU01b}. In the
present manuscript we report it for the case of Fe/W(110), Fe on bcc(110) solid solutions, and as
a function of the independently-controlled chemical nature of the interface. We also report the
dependence of the bimodal growth features upon growth conditions, island density, shape and ratio
of the two types. We have also tailored the growth of the flat islands to fabricate
nanometers-thick stripes that are self-organized by step-decoration at the atomic steps of the
buffer-layer. This is a fundamental advancement in self-organization as, up to now and to our
knowledge, only single- or double-AL stripes could be produced by
step-decoration\cite{bib-SHE97b,bib-HAU98b,bib-HAU98,bib-GAM02}. We already reported the growth of
such stripes on pure buffer layers\cite{bib-FRU02b,bib-FRU02c} and pure buffer layers covered with
ultrathin pseudomorphic interfacial layers\cite{bib-FRU04}. Here we also report the growth on
buffer layers of solid solution to get a better insight in the physical origin underlying the
bimodal growth. All these findings allow us to draw an enlarged panorama of growth processes of Fe
on cc(110) surfaces as a function of amount of material deposited, temperatures of growth and
possible annealing\bracketfigref{fig-growth-overview}. On \figref{fig-growth-overview} areas that
had not been investigated quantitatively in the literature prior to our reports are illustrated
with a picture of the resulting system. The reason why these areas had been previously overlooked
may come from the fact that studies on Fe(110) were already at the level of surface science when
the interest in self-assembly arose. Thus studies were undertaken by STM, restricting the study to
small lateral scales and amount of deposited material. Under these conditions it is not obvious to
recognize the bimodal feature of growth, which on the other side is straightforward with AFM on a
larger lateral scale. A discussion of the existing literature of the SK growth with respect to our
data on the bimodal growth is proposed at the end of the manuscript, with a view to identifying
whether the two types of islands had already been observed, however not identified as being of a
different nature. Finally, although we evidenced no discrepancy between any of our data and the
existing literature of Fe(110), concerning films or nanostructures, a difference of behavior
between PLD~(our technique) and MBE in the newly explored processes cannot be excluded.

\section{Compact Fe islands}
\label{sec-compact-dots}

\subsection{Overview}

\begin{figure}
\begin{center}
  \includegraphics[width=156mm]{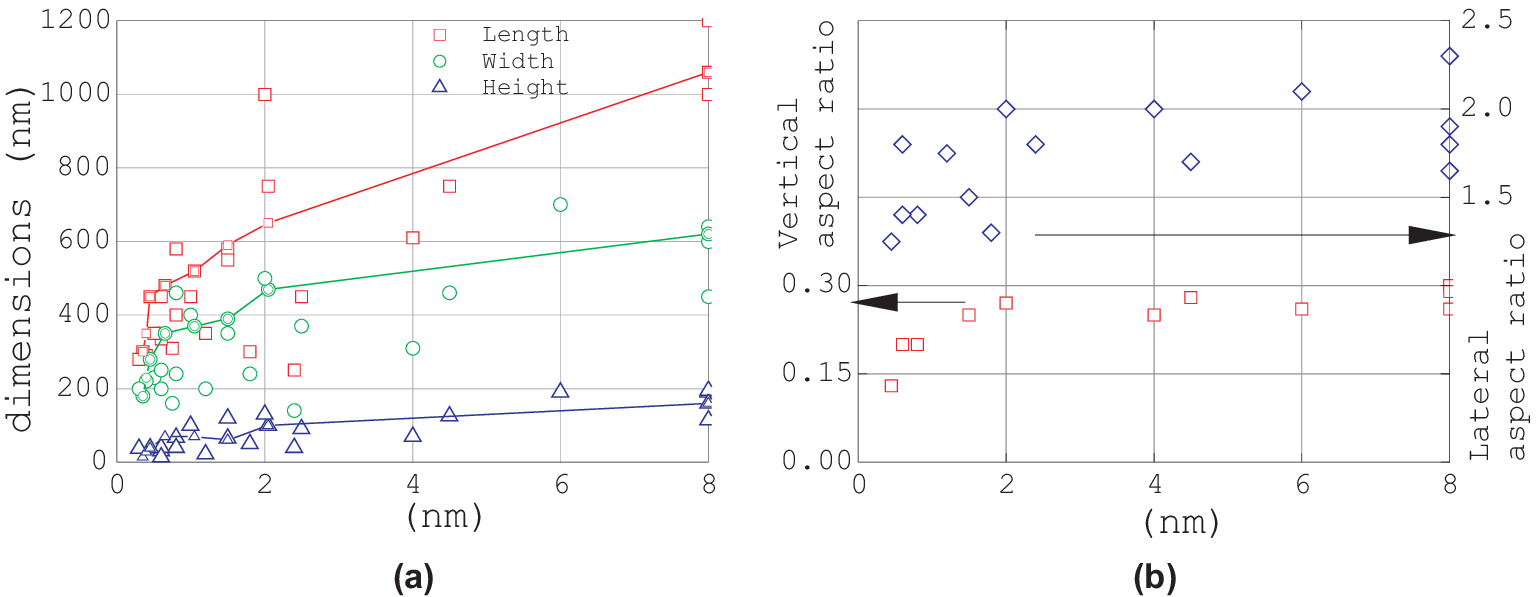}%
  \end{center}
  \caption{\label{fig-island-stat}\dataref{POJ-PhD-Ch2-Ilots.OPJ}(a)~Distribution and increase of the
  dimensions of compact dots for a series of samples grown at
  $\unit[450]{^\circ C}$, versus nominal thickness~$\Theta$. The full lines stand for one selected sample
  (b)~Vertical~(squares) and lateral~(diamonds) aspect ratio depending on growth temperature.}
\end{figure}

In the approximate range of temperature $\tempC{325-575}$ compact Fe islands are formed upon
deposition on Mo(110) and W(110)\bracketfigref{fig-bimodal-growth}. The areal density decreases
with temperature, from $\approx\unit[0.5]{\mu m^{-2}}$ at \tempC{325} to $\approx\unit[0.05]{\mu
m^{-2}}$ at \tempK{575} for a nominal thickness $\Theta\approx\thicknm{5}$. This density increases
very little in the range $\Theta=\thicknm{1-10}$, so that the mean dimensions of the dots scale
roughly with $\sqrt[3]{\Theta}$, ending up in a range of length
$\thicknm{100}-\thickmicron{1}$\bracketfigref{fig-island-stat}. On this figure it can be seen that
the distribution of mean size is significant from one sample to another. This suggests that the
nucleation of dots does not depend solely on growth temperature but may also depend on extrinsic
parameters such as step density and orientation\cite{bib-FRU06b} or residual contaminants at the
surface.

\begin{figure}
\begin{center}
  \includegraphics[width=158mm]{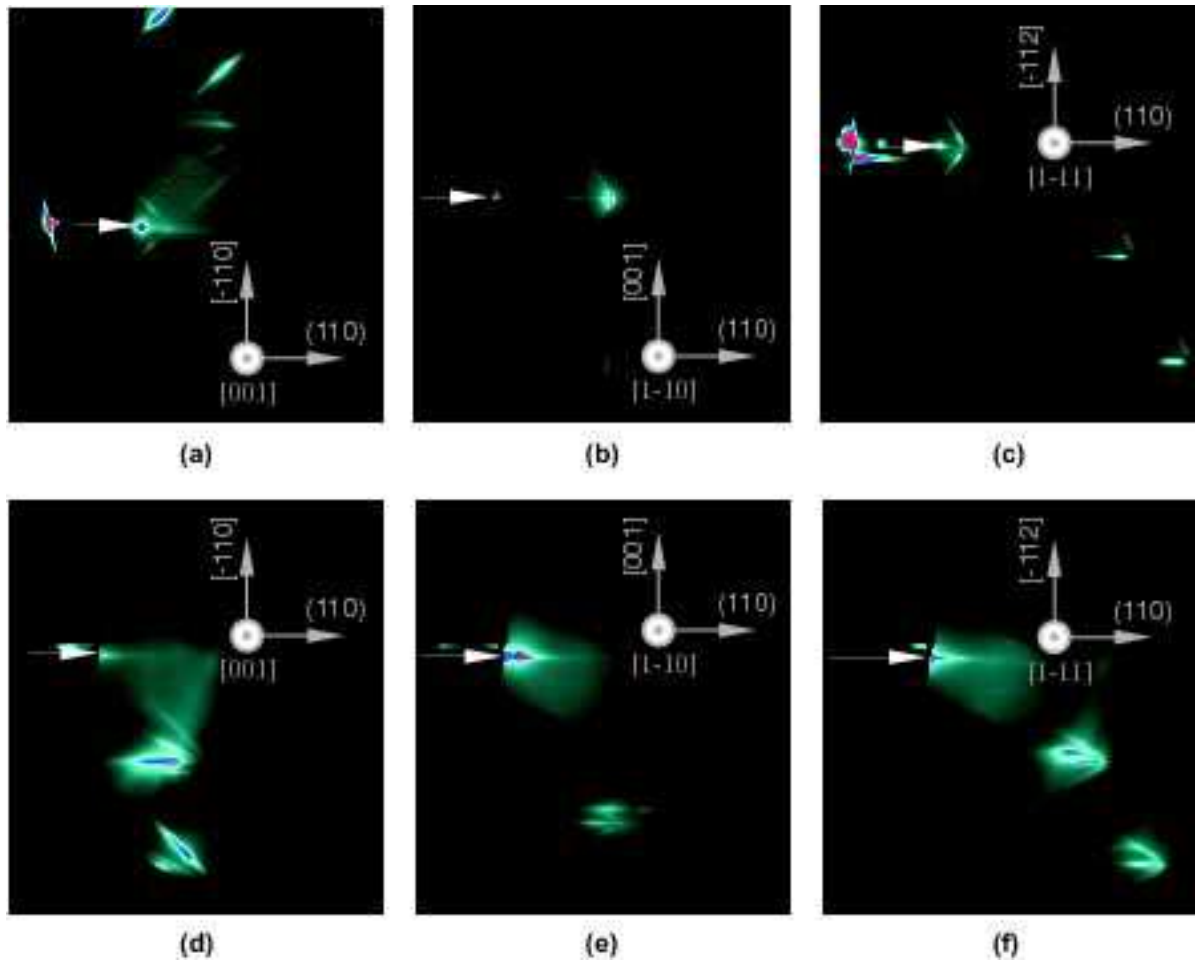}%
  \end{center}
  \caption{\label{fig-rheed-facets}\dataref{FRU136 pour Mo; FRU137 pour W}RHEED patterns of compact Fe(110)
  dots deposited on (a-c) Mo(110) and (d-f) W(110). The sample lies on the left of the patterns, with the mean surface vertical.
  The azimuths are (a,d)~$[001]$, (b,e)~$[1-10]$, (c,f)~$[1-11]$.
  The plane of incidence is indicated with a white arrow pointing on the reflected beam. The sample is rotated a few degrees off
  these azimuths for all images, because this allows to better evidence the tilted streaks arising from
  facets. The facets evidenced are of type (a)~$\{200\}$ and weaker $\{310\}$ (b)~$\{112\}$ and weaker $\{332\}$
  (c)~$\{110\}$ and weaker $\{211\}$ (d)~$\{200\}$ and $\{310\}$ (e)~$\{332\}$ (f)~$\{420\}$~(see Appendix~2).
  }
\end{figure}

A closeup view of the dots reveals that they are facetted, and elongated along
[001]\bracketsubfigref{fig-bimodal-growth}{b,d}. The smoothness at the atomic scale of the top
facet can be checked directly with AFM, with often no or only very few single atomic steps. No
emerging screw dislocation has ever been observed either by  this mean. The tilted facets are
probed by RHEED in the form of tilted diffraction streaks, each perpendicular to one set of
facets\bracketfigref{fig-rheed-facets}. The tilt angle of the facets is inferred  directly from
these patterns, much more reliably than with AFM were tip shading effects may arise for such steep
slopes. The sharpness of the streaks suggests the smoothness at the atomic scale of the facets.
The values of the angles of the facets are analyzed quantitatively in the following paragraph. On
\figref{fig-rheed-facets} notice also the doublets of horizontal streaks, revealing the in-plane
lattice parameters of both the substrate and Fe, arising from pseudomorphic areas and top facets
of Fe islands, respectively. The doublets are better seen for
Fe/W(110)\bracketsubfigref{fig-rheed-facets}{d-f}, where it is also clear that the intercept of
the tilted streaks occurs on the horizontal streaks associated with Fe, as expected. XRD was
performed in the $\theta-2\theta$ geometry, showing that the lattice parameter of Fe in the dots
equals that of bulk within better than $\unit[0.1]{\%}$. The crystalline coherence length,
extracted from the analysis of the peak width using the Scherrer formula, equals the mean
thickness of the dots. Owing first to the size distribution between dots, second to the
distribution of local heights in each dot resulting from the tilted facets, this peak does no
display side oscillations like for smooth films in \figref{fig-sapphire-kissing}.

\begin{figure}
  \begin{center}
  \includegraphics[width=128mm]{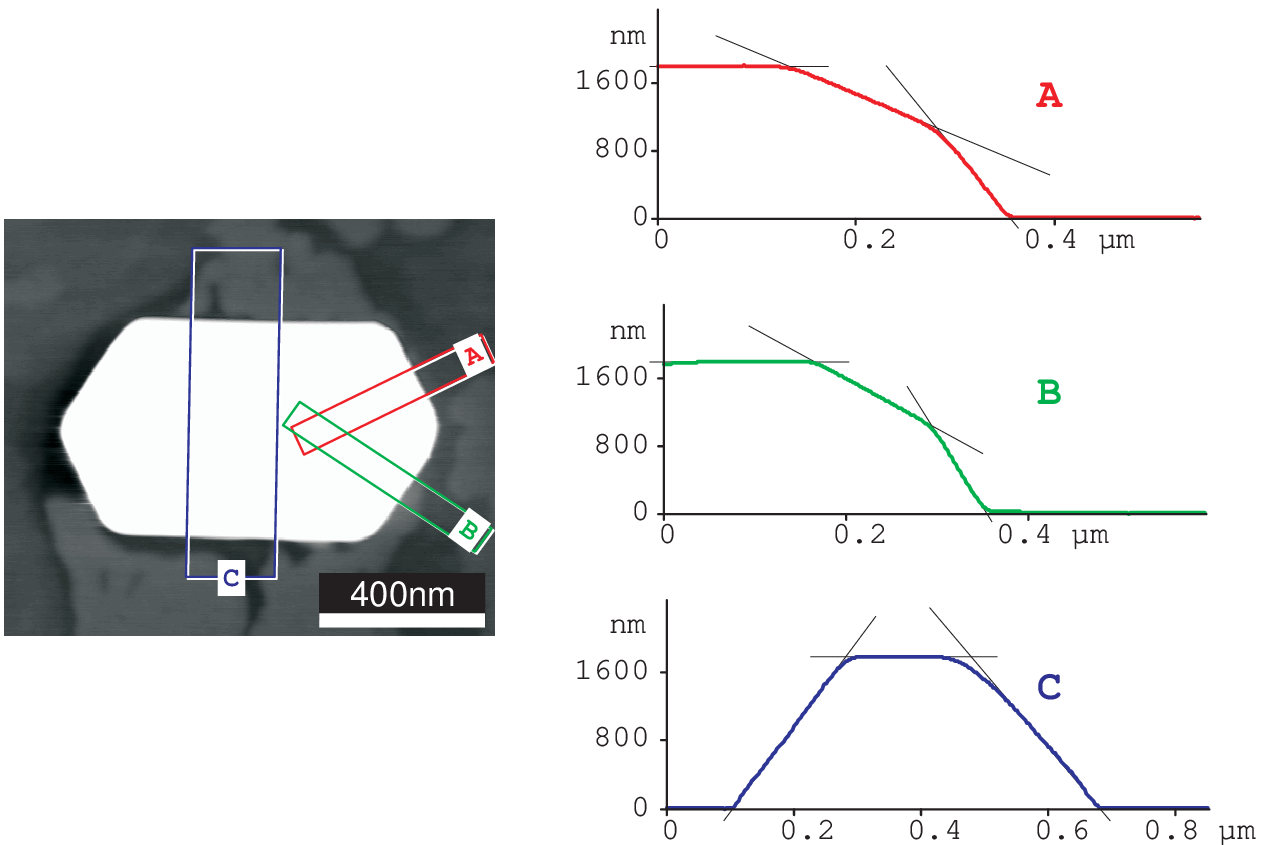}%
  \end{center}
  \caption{\label{fig-plot-cross-section}\dataref{Fru136 (0404s040.hdf)}Experimental cross-sections averaged laterally,
  performed on the dot shown in \subfigref{fig-bimodal-growth}b, labelled A, B, C on the AFM picture.
  Thin straight lines highlight the facets on the cross-sections.}
\end{figure}

\subsection{Wulff-Kaishev construction for Fe(110) dots}

\begin{figure}[t]
\begin{center}
{%
  \includegraphics[width=158mm]{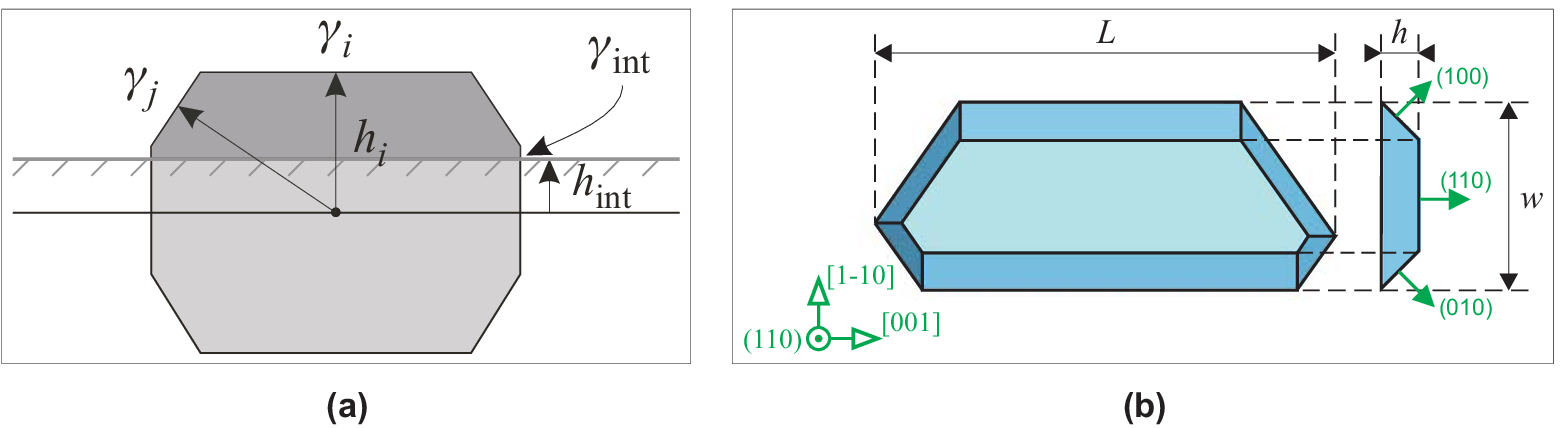}%
  \caption{(a)~Notations used for the Wulff-Kaishev theorem. The dashed surface stands for the
  supporting surface and the darker part of the crystal is the supported crystal.
  (b)~Definition of the full width~$w$, height~$h$ and full length~$L$, which can also be defined if
  the dot does not display symmetric facets, as illustrated here.} \label{fig-wulff-kaischev}}
\end{center}
\end{figure}

In the view of the above topographic and structural data, self-assembled Fe(110) dots can be
considered to be elastically relaxed single crystals. The shape of such crystals at equilibrium on
a supporting surface is described by the Wulff-Kaishev
construction\bracketsubfigref{fig-wulff-kaischev}a:

\begin{equation}
  \label{eq-WulffKaichev}\frac{\gamma_{i}}{h_{i}} = \frac{(\gammaint-\gamma_\mathrm{S})}{h_\mathrm{int}} =
  \mathrm{Constant}
\end{equation}

\begin{figure}
\begin{center}
  \includegraphics[width=160mm]{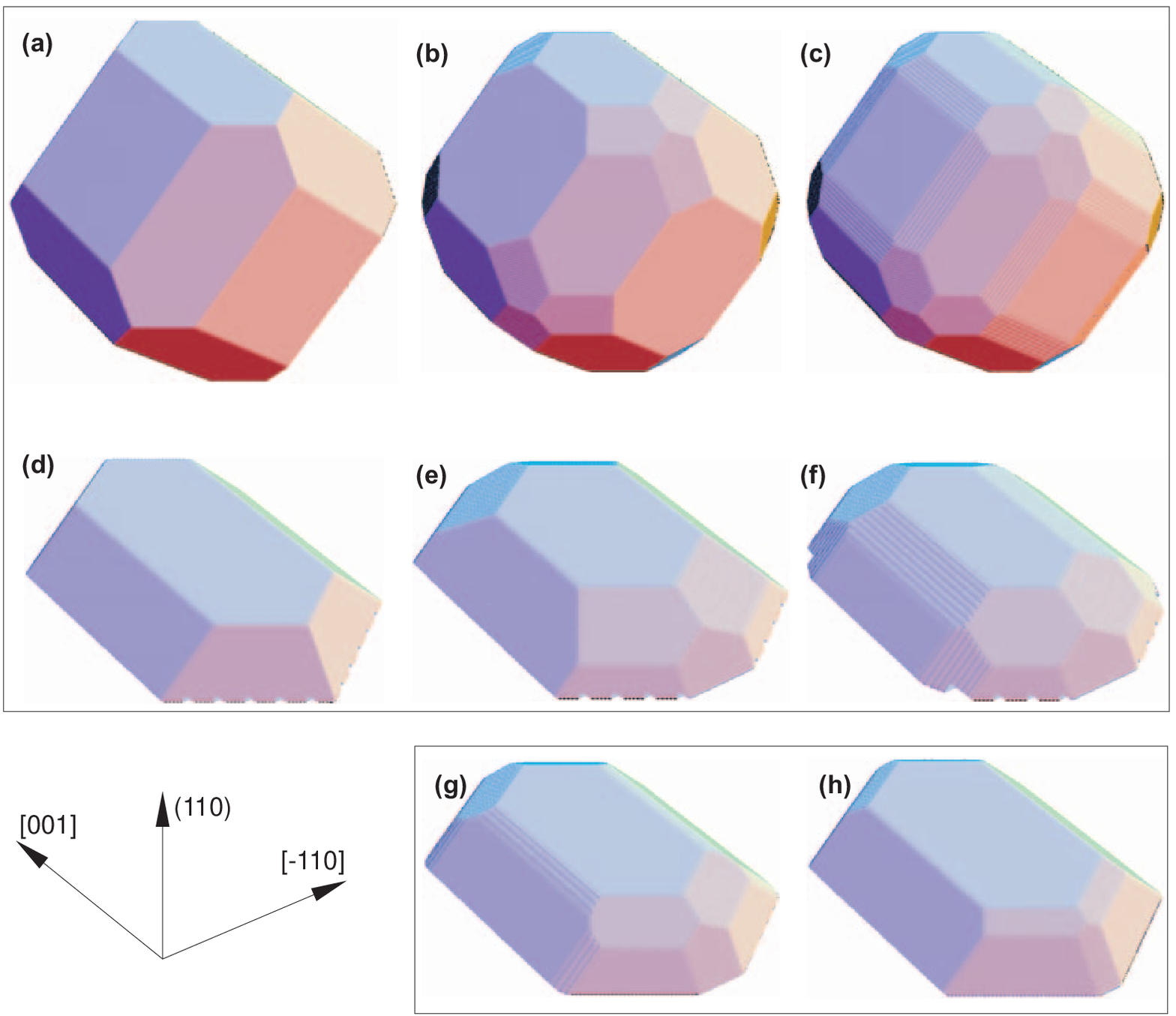}%
\end{center}
  \caption{\label{fig-all-facets}Wulff (a-c) and Wulff-Kaishev (d-h) constructions. (a-f) are drawn
  based on the values of surface energy computed by Vitos\cite{bib-VIT98}:
  $\gamma_{\{110\}}=\unit[2.430]{J/m^2}$;
  $\gamma_{\{001\}}=\unit[2.222]{J/m^2}$; $\gamma_{\{211\}}=\unit[2.589]{J/m^2}$;
  $\gamma_{\{310\}}=\unit[2.393]{J/m^2}$. For (g) and (h) $\gamma_{\{110\}}$, $\gamma_{\{001\}}$
  and $\gamma_{\mathrm{int}}$ have been reduced by $\unit[2.5]{\%}$ and $\unit[5]{\%}$,
  respectively.
  (d-h) are drawn based on the interfacial energy determined experimentally: $\gamma_{\mathrm{int}}=\unit[0.65]{J/m^2}$.
  The facets taken into account are $\{001\}$ and $\{110\}$ in (a,d), adding $\{211\}$ in (b,e)
  and $\{310\}$ in (c,f-h). $\{332\}$ facets are not shown because their energy was not reported by Vitos,
  although these facets are evidenced by RHEED in the
  experiments\bracketfigref{fig-bimodal-growth}.}
\end{figure}

where $\gamma_i$ is the free energy of facets~$i$ of a given family of crystallographic planes of
the material, $\gamma_\mathrm{S}$ is that of the free surface of the substrate, and $\gammaint$ is
the interfacial free energy\cite{bib-MUL00}. From the calculated surface energies\cite{bib-VIT98},
the free surfaces of a free-standing Fe crystal are expected to be mainly of type $\{110\}$ and
$\{100\}$, with secondary $\{211\}$ and $\{310\}$ facets. RHEED is consistent in revealing these
four types of facets, plus $\{332\}$ along the electron
azimuth~$<1-10>$\bracketfigref{fig-rheed-facets}. Notice however that not all dots display facets
other than $\{001\}$ and $\{110\}$. Finally, sharp edges are scarcely observed by AFM. It is not
clear whether this stems from the rounded shape of the tip at the scale of a few tens of
nanometers, or whether edges are really rounded.

\begin{figure}[t]
\begin{center}
  \includegraphics[width=103mm]{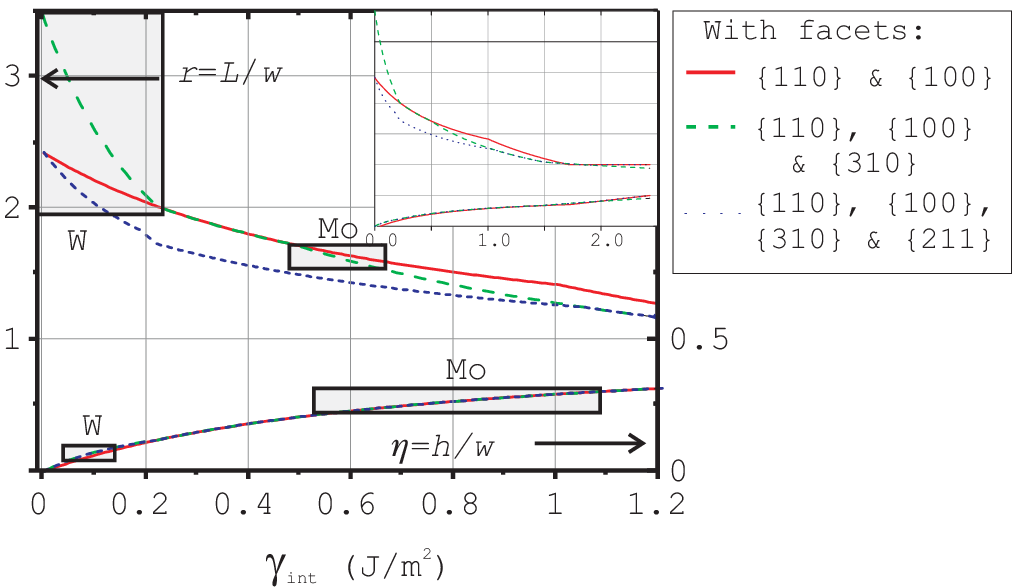}%
\end{center}
  \caption{\label{fig-plot-aspect-ratios}Lateral $r=L/w$ and vertical $\eta=h/l$ aspect ratios of
  supported Fe(110) dots depending on the interfacial energy $\gammaint$. The calculations were
  made considering two, three and four types of facets, and assuming $\gamma_\mathrm{S}=\gamma_{\{110\}}$~(see Appendix~I).
  The vertical size of the rectangles stands
  for the spread of experimental aspect ratios observed for Fe/W~(left) and Fe/Mo~(right).}
\end{figure}

The schematic shape of both types of facetted crystals are shown in \figref{fig-all-facets}. For
comparison notice in \figref{fig-rheed-facets} that for $[1-10]$ and $[1-11]$ azimuths facets with
the highest angle arise for Fe/Mo however not for Fe/W. This is explained by the fact that Fe/W
crystals are flatter than Fe/Mo crystals so that steep facets are not involved, see next paragraph
where the aspect ratios of the dots are analyzed quantitatively. For this analysis we introduce
two parameters to characterize the islands: the lateral aspect ratio $r=L/w$ and the vertical
aspect ratio $\eta=h/w$ where $L$, $w$ and $h$ are respectively the base full length, the base
full width and the height\bracketsubfigref{fig-wulff-kaischev}b. These aspect ratios can be
derived from \eqnref{eq-WulffKaichev}, see Appendix~I.  The calculated variation of $r$ and $\eta$
versus $\gammaint$ are shown on \figref{fig-plot-aspect-ratios}, assuming
$\gamma_\mathrm{S}=\gamma_{\{110\}}$, see the discussion at the end of this section. The variation
of $\eta$ does not depend on the number of facets considered excepted for very thin dots where
$\{310\}$ facets come into play~(see \figref{fig-all-facets}). On the reverse for any thickness
the length of the dot is reduced upon the consideration of additional facets, so that not all
curves for $r$ are superimposed.

\subsection{Experimental determination of interfacial energies}

The size and aspect ratios of dots on a given sample are distributed, see
\figref{fig-bimodal-growth}. The mean experimental values of $r$ and $\eta$ are nearly independent
on growth temperature~(not shown in this manuscript) and nominal
thickness\bracketfigref{fig-island-stat}, as expected for an equilibrium situation resulting
solely from the minimization of surface energy, with a vanishing elastic energy. We find
$\eta\simeq0.25$ and $r\simeq1.6$ for Fe/Mo, and $\eta\simeq0.1$ and $r\simeq2.7$ for Fe/W. These
values and their experimental distribution are reported on \figref{fig-plot-aspect-ratios} on the
$y$ axis, defining segments. The calculated curves of aspect ratios were used to convert these
$y$-segments in segments along the $x$ axis, defining the rectangles shown in the figure. This
determines the range of interfacial energy values compatible with the observations. The relevance
of the Wulff-Kaishev construction here is indicated by the overlap of the range of interfacial
energies compatible with both the observed lateral and vertical aspect ratios. We deduce
$\gamma_\mathrm{int,Mo}=\unit[0.65\pm0.15]{J.m^{-2}}$ and
$\gamma_\mathrm{int,W}=\unit[0.10\pm0.05]{J.m^{-2}}$. The most probable values were derived based
mostly on the value of $\eta$. Indeed, first the experimental distribution is found to be larger
for $r$ than for $\eta$, second the mean value of $r$ was found to vary slightly more then during
growth than that of $\eta$\bracketsubfigref{fig-island-stat}b. The experimental mean values for
$r$ and $\eta$ have also been measured for combinations of buffer layer and pseudomorphic
interfacial layers, \ie Mo/W and W/Mo. It was found that $\gammaint$ is related to the chemical
nature of the interface or interfacial layer, not on the in-plane lattice parameter imposed by the
underlaying buffer layer. These facts are discussed later. In the future these studies should be
extended to CGL buffer with Mo or W interfacial layers, so that the contribution of dislocations
and elastic energy on effective interfacial energies is derived.

\subsection{Limits of the temperature range for the growth of compact dots}

The in-plane shape of the dots becomes more and more irregular when the growth temperature is
reduced below $\tempC{\approx350}$, and the tilted facets tend to disappear. At the opposite
range, when the growth temperature is raised above $\tempC{\approx500}$ the tilted facets are
still well defined, but the aspect ratios become more and more distributed. The distribution of
$\eta$ broadens less than that of $r$. Thus, the optimum range of temperatures for growing faceted
dots with a moderate distribution of aspect ratios is $\tempC{350}-\tempC{500}$. It will be shown
in \secref{sec-flat-islands} for Fe/Mo that at still higher temperatures the SK growth is
initiated with flat dots and not with compact dots.

\subsection{Changes of shape induced by surfactant-mediated growth}

Surfactants are used in epitaxial growth to promote a layer-by-layer growth mode like with Pb with
Co/Cu(111)\cite{bib-CAM96} or O for Fe(001)\cite{bib-BIS99}. Kinetic effects are often identified
as the leading effect promoting layer-by-layer growth\cite{bib-FER00}, by lowering diffusion
energies. A side effect is the lowering of surface energies when surfactants are present. The
ratio of the different surface energies might be also changed, which should a modification of the
shape of crystals.

\begin{figure}
\begin{center}
  \includegraphics[width=134mm]{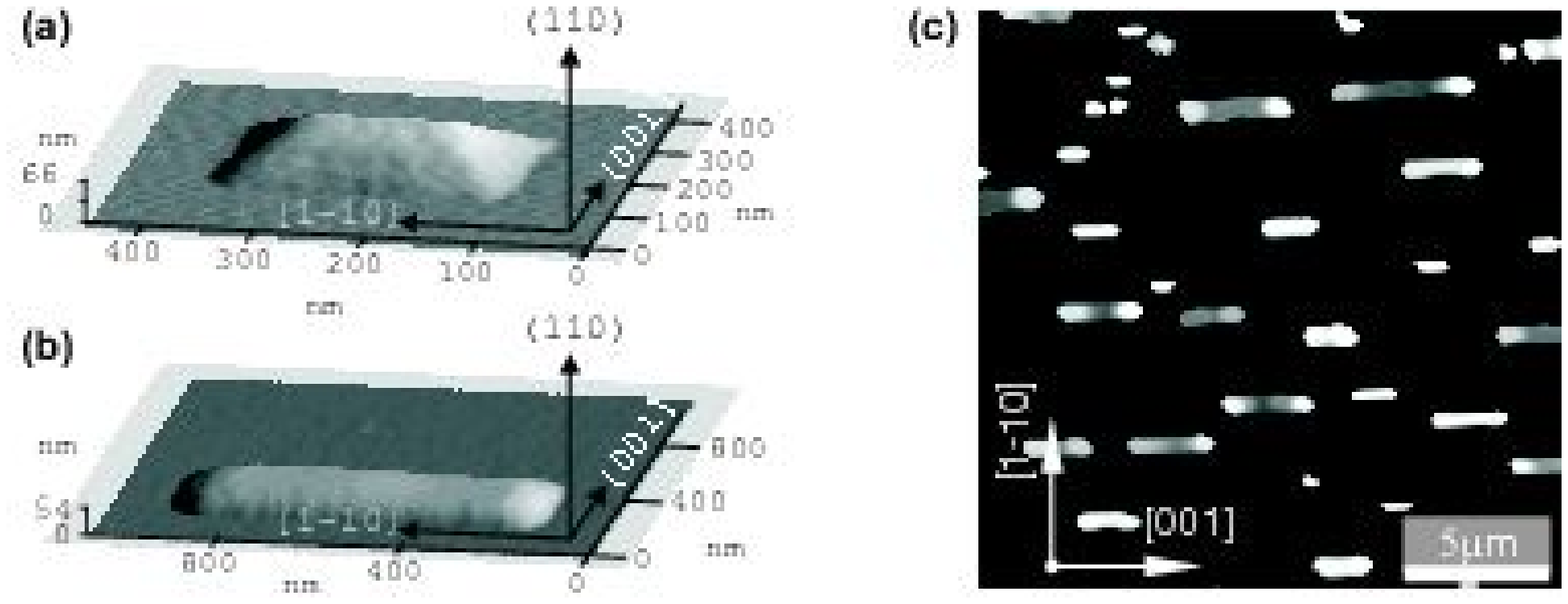}%
\end{center}
  \caption{\label{fig-surfactants} AFM images of an Fe(110) dot obtained
  \dataref{POJ/PO14/CTND/1102Z004.HDF}(a)~upon co-deposition of Fe and Sm at \tempC{450}
  \dataref{POJ/PO108/1215d004.HDF}(b)~upon deposition at \tempC{475} of Fe on
  \thickAL{1} Sm/Mo(110)
  (b)~\dataref{FC13/0422T01D.hdf}$\unit[20\times20]{\mu\mathrm{m}}$ AFM image of Fe(110) dots
  deposited under poor vacuum~($\sim\unit[\scientific{1}{-9}]{Torr}$).}
\end{figure}

We have observed a surfactant effect of Sm on Fe(110) dots. Upon deposition from a
$\mathrm{SmFe}_2$ compound target, we have observed in a narrow range of temperatures around
$\tempC{450}$ the formation of hut-shaped dots\bracketsubfigref{fig-surfactants}a\cite{bib-JUB01}.
XRD revealed that the dots are made of pure Fe relaxed to the bulk lattice parameter with no
indication of Sm or any SmFe compounds. This can be explained by the low evaporation temperature
of Sm under UHV, $\approx\tempC{200}$, so that Sm re-evaporates from the surface and does not
contribute to the formation of SmFe compounds, for which the present temperatures are too low.
Auger spectra revealed that a weak amount of Sm remains at the surface, probably one atomic layer,
that significantly lowers the surface energies of Fe and therefore whose evaporation is
hindered\cite{bib-STE87}. These hut-shaped dots were similarly obtained after the deposition of
pure Fe on $\sim\thickAL{1}$  Sm on Mo(110) at \tempC{475}\bracketsubfigref{fig-surfactants}b,
supporting the aforementioned growth mechanism using a $\mathrm{SmFe}_2$ target and confirming the
surfactant effect of Sm.

It is obvious that the surface energies of the various bcc-Fe facets are not decreased in a
proportional way, as the huts are elongated along $[1\overline10]$ and display $\{111\}$ facets on
the sides and $\{001\}$ facets at both ends. No $\{110\}$ facets are visible . A na{\"\i}ve explanation
why $\{111\}$ facets are favored to $\{110\}$ come from hexagonal symmetry of both Mo$\{111\}$ and
Sm $\{0001\}$, and also the lattice mismatch being smaller for Sm(0001)/Mo(111) than for
Sm(0001)/Mo(110), respectively \unit[12]{\%} versus \unit[15]{\%}. The real picture is more
complicated as the valence of Sm depends sensitively on strain and on the number of ALs
deposited\cite{bib-STE87,bib-STE87b,bib-STE89,bib-STE89b}.

Let us add a word about surface contamination by residual gases. In the rare cases where the level
of vacuum was poor during deposition, \ie typically $\unit[\scientific{1}{-9}]{Torr}$, the shape
of the dots was altered: elongated dots were formed, with their height significantly increased at
both ends in a shape of dome\bracketsubfigref{fig-surfactants}c. As these domes remain located at
the end of the dots, whose length increases during the course of growth, it is likely that they
move along the dot while growth proceeds. The composition of the surface contaminants has not been
analyzed with AES.

\subsection{Discussion}

This discussion mainly considers the case of Fe(110) dots without surfactants. Let us first stress
two points. First in our approach $\gammaint$ is a phenomenological value that takes into account
both the chemical hybridization at the interface, and the elastic energy of the array of
interfacial dislocations. This view is valid as soon as one considers islands that are
significantly higher than the lateral distance between dislocations, so that the resulting strain
field is averaged out in most of the island, except close to the interface. Second, the chemical
part of $\gammaint$ shall not be interpreted as the wetting energy of Fe on either W or Mo, but as
the wetting energy of Fe on either Fe(\thickAL1)/W or Fe(\thickAL1)/Mo. Indeed, the question is
whether the Fe atoms will form a smooth film, or self-assemble in islands, \emph{above} the
wetting \AL. Notice that this explains why surfactants like Sm are not efficient in suppressing
the SK growth of dislocated dots: both the surface energy of Fe and that of the 1AL-covered
substrate are reduced by the surfactant.

Let us now discuss the values determined for $\gammaint$ from the mean aspect ratio of the dots.
We first discuss the soundness of our analysis, which is based on the numerical values of surface
energy computed by Vitos \etal\cite{bib-VIT98}. The AFM views in \figref{fig-bimodal-growth}(b,d)
are in reasonable agreement with the theoretical picture of \subfigref{fig-all-facets}{d-f}, which
shows that the computed values of surface energy are rather accurate. However a statistical
analysis of the experimental dots with AFM however reveals several details. The facets other than
$\{001\}$ and $\{110\}$ are of smaller extent in the experiments, especially concerning $\{310\}$.
\subfigref{fig-all-facets}{g-h} show the shape of dots when $\gamma_{\{110\}}$, $\gamma_{\{001\}}$
and $\gamma_{\mathrm{int}}$ are all decreased by $\unit[2.5]{\%}$ and $\unit[5]{\%}$,
respectively. Theses shapes fit better the experiments. Thus, the surface energy of the facets of
higher Miller index might have been underestimated in relative value by a few percent in the
computation. More precise values cannot be extracted from our data because of the distribution of
shape from one dot to another. For instance, not all dots display the additional facets. Also
$\{211\}$ facets at the base of the dots (at the end of the arrows) are more rarely found than
those around the $\{110\}$ top. These uncertainties over the values of surface energy might
contribute to the only partial overlap of the ranges of interfacial energies compatible with the
experimental observations of $\eta$ and $r$, see \figref{fig-plot-aspect-ratios}

Let us now discuss the values derived for $\gammaint$. $\gammaint$ was found to depend on the
interfacial layer and not on the underlying buffer layer, which is not surprising. Indeed W and Mo
have similar lattice parameters, $\thickAA{3.165}$ and $\thickAA{3.147}$, respectively. Thus a
similar lattice misfit with Fe is expected, inducing a similar density of the interfacial
dislocations network, and thus a similar areal density of elastic energy. Concerning the
dependence on the interfacial layer, two phenomena may play a role. First the strain field of the
interfacial dislocations extends down mostly into this layer, not in the underlying buffer layer.
Thus, from the higher elastic constants of W over Mo one would expect
$\gamma_\mathrm{int,W}>\gamma_\mathrm{int,Mo}$. As the reverse relationship is observed the second
effect must therefore be dominant and with an opposite trend: the surface band structure of one
atomic layer of Fe has not recovered that of the surface of a bulk Fe crystal, and is still
influenced by the underlying material. More specifically, \eqnref{eq-WulffKaichev} shows that what
matters is not solely $\gammaint$, but $\gammaint-\gamma_\mathrm{S}$, $\mathrm{S}$ being here
respective to the wetting monolayer. We already mentioned that the plots of
\figref{fig-plot-aspect-ratios} has been computed assuming assuming
$\gamma_\mathrm{S}=\gamma_{\{110\}}$. This was assumed because no figure is known for the surface
energy of the monolayer of Fe(110). As a consequence it is not possible to disentangle $\gammaint$
from $\gamma_\mathrm{S}$. A point of view would be to conclude that Fe(\thickAL1)/W has a higher
surface energy than Fe(\thickAL1)/Mo, which seems reasonable as the same relationship holds for
the bulk materials: $\gamma_\mathrm{W}\sim\unit[4.0]{J/m^{2}}$ and
$\gamma_\mathrm{Mo}\sim\unit[3.4]{J/m^{2}}$.

\begin{figure}
\begin{center}
  \includegraphics[width=63mm]{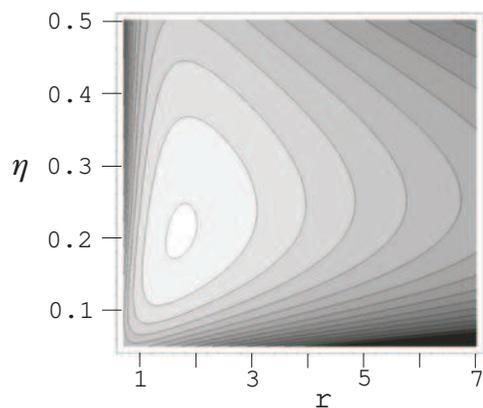}%
\end{center}
  \caption{\label{fig-calculated-energy-versus-shape}\dataref{PRB2001}Total
  surface/interface energy of a Fe(110) dot
  calculated with $\gammaint=\unit[0.5]{J/m^2}$. The minimum of energy is reached for $r=1.7$ and
  $\eta=0.2$ which corresponds to $\Es/V_0^{2/3}=\unit[4.33]{J/m^2}$. The energy difference between lines
  of equal energy is $\unit[200]{mJ/m^2}$.}
\end{figure}

So far we have neglected a potential residual strain in the dots. The residual strain is less than
0.1\% as probed by XRD, which corresponds to a stored elastic energy
$E_\mathrm{elas}/V_0^{2/3}=\unit[0.2]{J/m^2}$. Only a small fraction of $E_\mathrm{elas}$ may be
gained by changing the island shape, say approximately at most 10-20\% due to a moderate vertical
aspect ratio, which amounts to about$\unit[0.04]{J/m^2}$. As the equilibrium dot shape results
from the interplay of the elastic energy and the surface/interface energy, let us estimate how a
surface/interface energy excess of $\unit[0.04]{J/m^2}$ would influence the shape of the dot.
\figref{fig-calculated-energy-versus-shape} shows the total surface/interface energy per unit of
the island volume (more precisely the scalable quantity $\Es/V_0^{2/3}$) as a function of $\eta$
and $r$, plotted for $\gammaint=\unit[0.5]{J/m^2}$. The energy minimum is obtained for $r=1.7$ and
$\eta=0.2$ which corresponds to $\Es/V_0^{2/3}=\unit[4.33]{J/m^2}$. An excess of
$\unit[0.04]{J/m^2}$ in the surface energy would correspond to a variation in the aspect ratios of
$\Delta r=\pm0.4$ or $\Delta\eta=\pm0.05$. These deviations from the calculated equilibrium shape
are small and comparable to the experimental distributions. Thus the effect of residual elastic
strain can not be distinguished, at least directly. However strain effects might also contribute
for the only partial overlap of the ranges of interfacial energies compatible with the
experimental observations of $\eta$ and $r$, see \figref{fig-plot-aspect-ratios}.

\section{Flat islands and stripes}
\label{sec-flat-islands}

In this section we report results pertaining to flat nanostructures that arise in the bimodal
growth mode\bracketsubfigref{fig-bimodal-growth}a. Contrary to the case of compact islands that
grow homothetically in the course of deposition--see previous section, the height of the flat
nanostructures remains essentially constant during growth, while only their laterals dimensions
increase. The occurrence of one or several such heights, constant during growth and therefore
presumed to be stable or metastable, has been reported for several systems:
Ag/GaAs(110)\cite{bib-SMI96}, Ag/Si(111)\cite{bib-GAV99,bib-HUA98}, Ag/Fe(001)\cite{bib-LUH01},
Pb/Si(111)\cite{bib-ALT97,bib-YEH00,bib-SU01}, Pb/Cu(111)\cite{bib-OTE02}. These heights are
sometimes called magic height, because their occurrence is unexpected within the scheme of
classical growth modes, either Volmer-Weber~(VW), Franck van den Merwe~(FM) or
Stranski-Krastanov~(SK). In the following we will refer to these nanostructures as
\emph{preferred} height instead of \emph{magic} height. We first give evidence for the bimodal SK
growth mode for Fe(110), then demonstrate the metastability of the preferred height and discuss
possible physical phenomena that can give rise to such effects. Finally we show how the occurrence
of a preferred height can be used to fabricate unusual self-organized systems, namely thick
stripes by step decoration, and nano-ties.

\subsection{Bimodal SK growth}

\begin{figure}
\begin{center}
  \includegraphics[width=149mm]{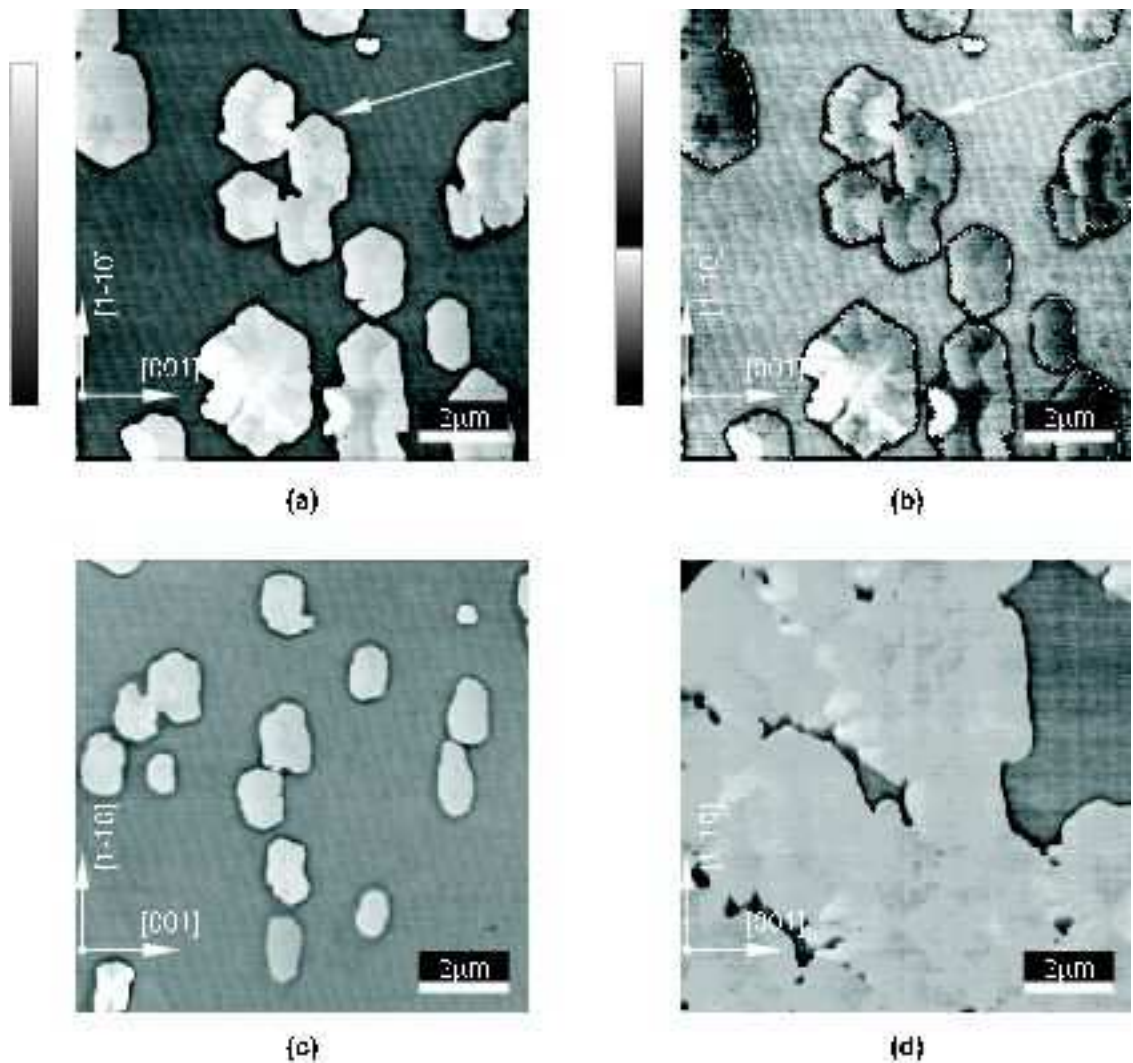}%
\end{center}
  \caption{\label{fig-wedged-dots}\dataref{FRU83/0608B024.hdf}Flat Fe(110) islands resulting from the
  deposition at \tempC{800} on Mo, seen on $\thickmicron{10\times10}$ AFM images.
  The white arrows indicate the downward direction of the array
  of mono atomic steps of the buffer layer (a)~$\Theta=\thicknm{0.5}$ normal scale~(grey scale shown
  on the left) (b)~same picture with a split
   scale~(see second scale bar) to achieve a significant
  contrast for both the background and the top of the islands.
  (c)~\dataref{0608A02A.BMP}$\Theta=\thicknm{0.25}$
  (d)~\dataref{06080002.BMP}$\Theta=\thicknm{1.25}$}
\end{figure}

\begin{figure}
\begin{center}
  \includegraphics[width=130mm]{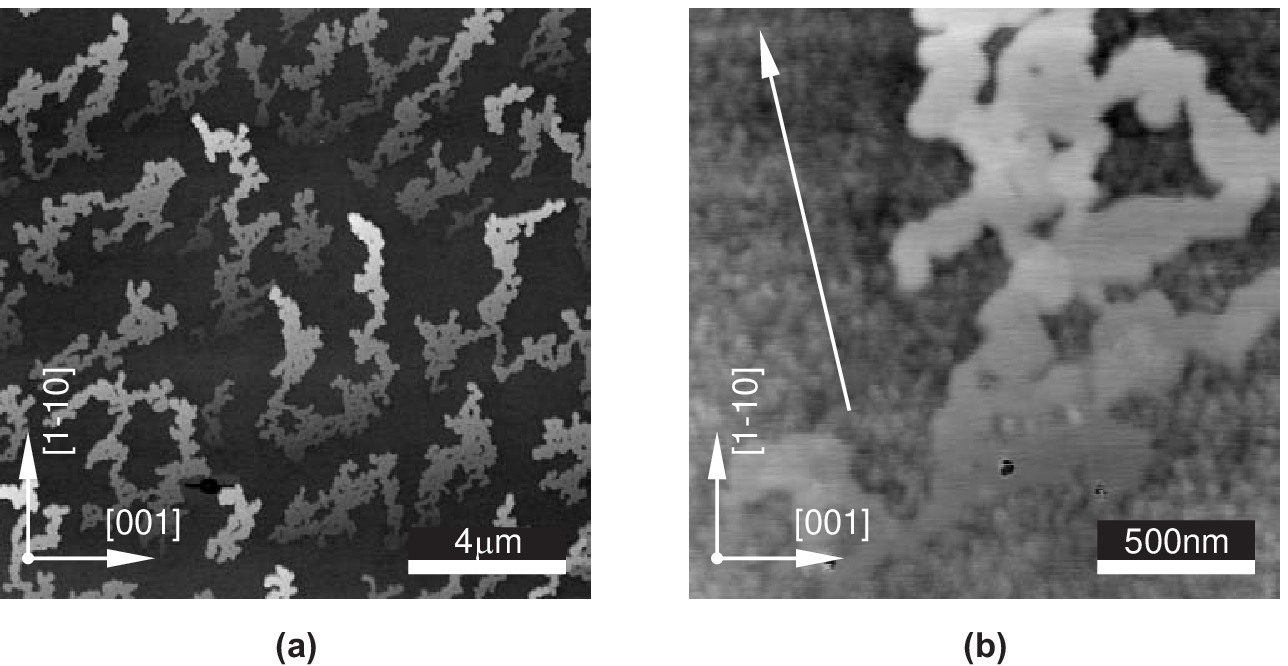}%
\end{center}
  \caption{\label{fig-au-on-mo}\dataref{(a)~FRU82/1023s000.hdf (b)~FRU82/1023s00c.hdf}AFM images
  of Au($\Theta=\thicknm{7.5}$)/Mo(110) upon deposition at $\tempC{600}$. The white arrow in (b)~indicates the downward direction of the array
  of mono atomic steps of the buffer layer.}
\end{figure}

Most of our experiments were performed on Mo(110) or Mo-covered W(110). We evidenced the bimodal
growth of Fe(110) in the approximate range of temperature $\tempC{250\sim500}$ (the bounds were
not determined accurately), as shown in \subfigref{fig-bimodal-growth}a. Apart from the compact
islands appearing as bright dots in this figure, and the wetting layer appearing as the darkest
background with an array of parallel monoatomic steps, flat islands with a large ratio of lateral
size over height are observed. Important to notice is that the height of these islands essentially
does not vary during growth. The further deposition of material only contributes to increase their
area through lateral growth. For deposition on pure Mo buffer layers the distribution of heights
is narrow: $h=\thicknm{1.2\pm0.2}$ above the wetting AL, \ie implying a mean total height of
\thickAL{7} above Mo. Details about the size and density of these islands can be found in
Ref.~\cite{bib-FRU01b}. Their shape depends on the growth temperature, see
Figs.~\ref{fig-bimodal-growth}a, \ref{fig-wedged-dots}, \ref{fig-nanobowties}. For the lowest
range of temperatures for the bimodal SK growth, \ie around \tempC{400}, the islands seem to grow
in the direction perpendicular to the steps of the substrate, with a shape influenced by single
atomic steps in a saw-tooth-like manner, see \figref{fig-bimodal-growth}. At higher temperatures
the islands have a tendency to display a hexagonal shape elongated parallel to $[001]$, be it
while the growth remains bimodal~(not shown here) or when only wedged islands grow at very high
temperature\bracketfigref{fig-wedged-dots}. In all cases notice that the top of these
nanostructures is not perfectly flat, but displays a weak density of single atomic steps to
compensate for the residual miscut, thereby allowing to keep a mean constant thickness over
lateral distances over micronmeters\bracketfigref{fig-wedged-dots}. This confirms the stability of
the preferred height. If it were not stable, these steps would flow over the top during growth,
ending up in wedged islands. This phenomenon is observed in the first stages of growth of
Fe/Mo(110)\cite{bib-MUR02} and Fe/W(110)\cite{bib-BET95}, below the constant preferred height.
When no preferred height exists, like for Au/Mo(110), these islands grow endlessly laterally in
the downward step direction\bracketfigref{fig-au-on-mo}. Beyond the case of Mo(110), the formation
of wedged islands is a common phenomenon on surfaces displaying atomic
steps\cite{bib-BET95,bib-ALT97,bib-REP02b} and the microscopic phenomena explaining the growth or
diffusion along the downward direction have been unveiled by monitoring growth
\insitu\cite{bib-LIN04}. Coming back to Fe(110), deposition on W is qualitatively similar to the
case of Mo, however with a higher preferred height and a broader distribution,
$t=\thicknm{6\pm2}$. Notice also that the density~(resp. the lateral size) of these islands is
lower~(resp. larger) than for deposition on~Mo.

\subsection{Evidence for the metastability of the preferred height}

The occurrence of flat islands whose height does not vary during growth and is independent from
the deposition temperature, suggests that their exists a stable or metastable height for
Fe/bcc(110) systems. It is the purpose of this sub-section to support this assumption. The
possible physical driving forces for the occurrence of such a preferred height will be discussed
in sub-section \ref{subsec-origin-magical-height}.

\begin{figure}
\begin{center}
  \includegraphics[width=158mm]{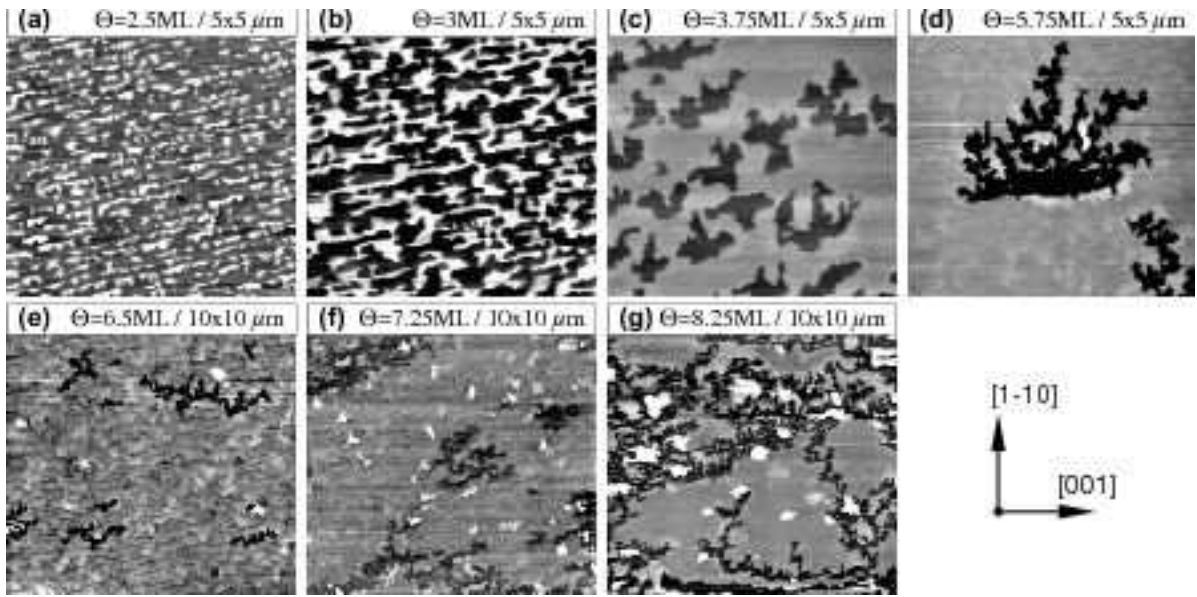}%
\end{center}
  \caption{\label{fig-metastable-thickness}\dataref{HDR, fig.I.24}AFM images
  of Fe deposited on Mo(110) at room temperature, followed by annealing at \tempC{450}, for several
  nominal thicknesses~$\Theta$. Notice the different scale
  used for the two rows.}
\end{figure}

Fe was deposited at RT on Mo(110) with a nominal thickness $\Theta=\thicknm{0-2}$ varied as a
wedge over the entire length of the substrate. The growth resulted in a roof-like rough
surface\cite{bib-FRU98b}, as first reported in Ref.\cite{bib-ALB93}, and similar to the rough
surfaces of moderate-temperature-deposited Mo and W layers reported in
\secref{sec-buffer}\bracketfigref{fig-rough-one}. Upon annealing at $\tempC{450}$ this rough film
unwets Mo, and thus Fe nanostructures are formed\bracketfigref{fig-metastable-thickness}. For
$\Theta\leq\thickAL{7}$ the nanostructures have all the same uniform height
$t=\thicknm{1.2\pm0.2}$. The area covered by the nanostructures increases as a function of nominal
thickness, until a nearly continuous layer $\thickAL{7}$-high is formed. We hereby confirm that
the occurrence of a preferred height gives the mean to achieve very smooth continuous films for SK
systems. From this experiment we also conclude that $\thickAL{7}$ is a highly preferred height for
Fe/Mo(110). For $\Theta>\thickAL{7}$ compact islands of height much larger than $\thickAL{7}$
nucleate, and for most of them grow at the expense of the continuous film: channels have formed in
this film, an effect which therefore seems to result from the diffusion of atoms towards the
compact islands. From the latter observation we finally conclude that $\thickAL{7}$ is a
\emph{metastable} thickness, \ie it is favorable against other numbers of ALs in the range of
small thickness, however it is unstable with respect to much thicker structures.

The experiment of room temperature deposition followed by annealing has not been conducted for
Fe/W(110). However, we will see that the experiment on self-organized stripes, reported hereafter,
leads us to the conclusion that a metastable thickness also exists for this system.

\subsection{Fabrication of wires self-organized along steps}
\label{subsec-wires}

In this sub-section we show how the (meta)stability of layers with a preferred height can be used
to produce self-organized thick stripes decorating the array of mono-atomic steps of buffer
layers. By \emph{thick} we mean much higher than one or two atomic layer(s), which are the heights
usually reported in the literature, for self-organized stripes fabricated by step decoration.

\begin{figure}
\begin{center}
  \includegraphics[width=150mm]{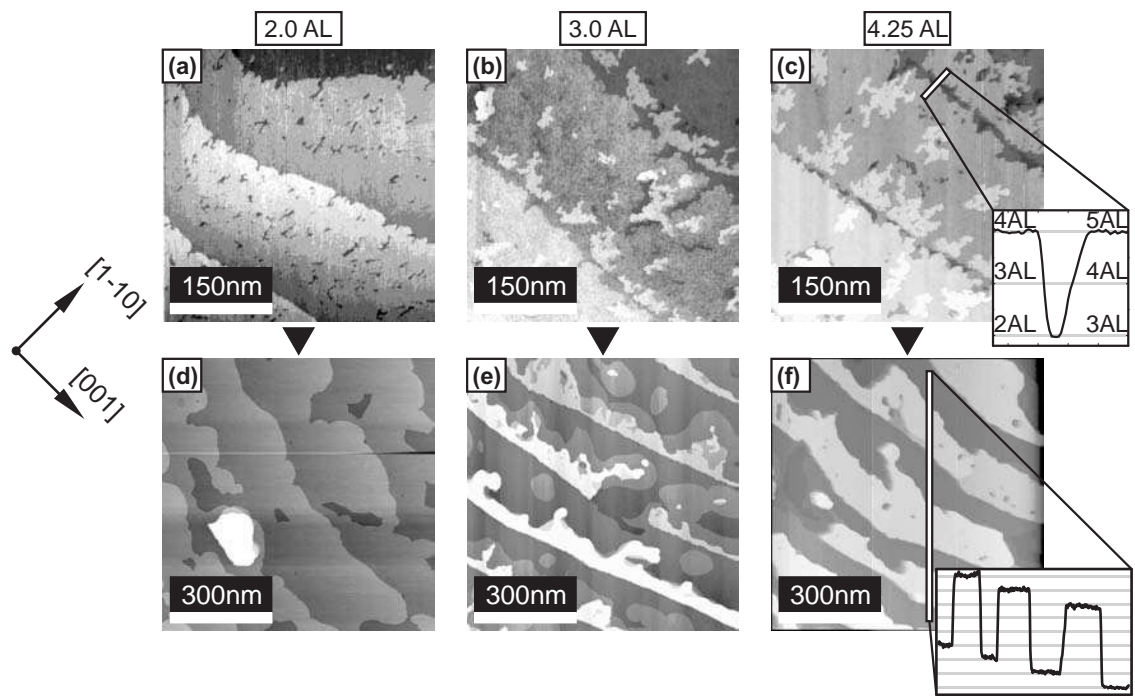}%
\end{center}
  \caption{\label{fig-stripes-with-stm}\dataref{From APL 2004 and DHDR}STM images of Fe/Mo(110)
  (a-c)~after deposition at $\tempC{200}$ and (d-f)~after annealing at $\tempC{450}$.}
\end{figure}

\begin{figure}
\begin{center}
  \includegraphics[width=157mm]{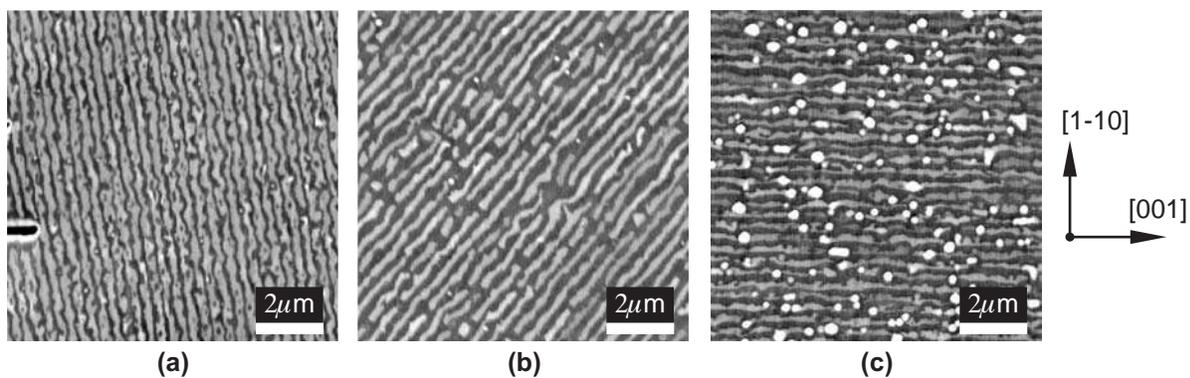}%
\end{center}
  \caption{\label{fig-stripes-various-orientations}\dataref{Fig HDR, from PhD POJ}AFM images
  of stripes formed by deposition of Fe on Mo(110) at $\tempC{200}$, followed by annealing at $\tempC{450}$.
  The direction of the stripes follows that of the array of atomic steps of the buffer layer, itself
  in turn dictated by the miscut of the substrate. In (c) a non-optimized growth procedure (slightly too high temperature)
  is responsible for the occurrence of somewhat thicker dots in the course of the stripes. }
\end{figure}

On \subfigref{fig-metastable-thickness}{a-b} the nanostructures display a preferential alignment
along one single well-defined direction. This direction is not a low Miller index crystalline
axis, but is related to the underlying array of mono-atomic steps of the buffer layer. Indeed,
comparison with AFM images of the raw buffer layers reveals that these stripe-like features
display the same orientation, and the same period across the stripes, as that of the array of
steps on the Mo surface. These steps are thus expected to play a role in the unwetting process.
The tendency to form stripes can be enhanced if deposition is performed slightly above room
temperature, in the range around $\tempC{150-200}$, followed again by annealing around
$\tempC{450}$, the exact value of the latter temperature being not crucial in the
process\bracketfigref{fig-stripes-various-orientations}. The reason for improvement is revealed by
STM images taken before and after annealing\bracketfigref{fig-stripes-with-stm}\cite{bib-FRU04}.
Growth at $\tempC{200}$ proceeds layer-by-layer. However beyond $\thickAL2$ grooves appear at the
surface of the film, with the same orientation and spacing as the array of mono-atomic step of Mo.
It is likely that the grooves appear above these buried mono-atomic steps, driven by the strain
resulting from the misfit of height between (110) atomic layers of Mo and Fe, $\thickAA{2.23}$ and
$\thickAA{2.03}$, respectively. The tendency to have grooves close to steps of the buffer layer
could be seen on a Fig.1b of Ref.\cite{bib-MUR02} for Fe grown on Mo at RT. The grooves then
probably act as nucleation sites for unwetting during the annealing, yielding stripes. Notice that
the straighter side of the stripes lies along the ascending step, probably because the strain acts
as a barrier. It has not been attempted to systematically investigate annealing or other
procedures to improve the free side of the stripes, which is much rougher on this image. We
however stress that neither deposition nor annealing temperatures are critical parameters for
getting stripes. Notice that such stripes can be fabricated in any orientation away from $[001]$
to $[1\overline10]$\bracketfigref{fig-stripes-various-orientations}, which definitely shows that
there is no correlation between their orientation and crystallographic directions. Notice that in
our early reports non-connected Fe/Mo(110) stripes, of irregular shape and height, were produced
by deposition at $\tempC{250}$ followed by annealing at \tempC{450}\cite{bib-FRU02b}.

\begin{figure}
\begin{center}
  \includegraphics[width=89mm]{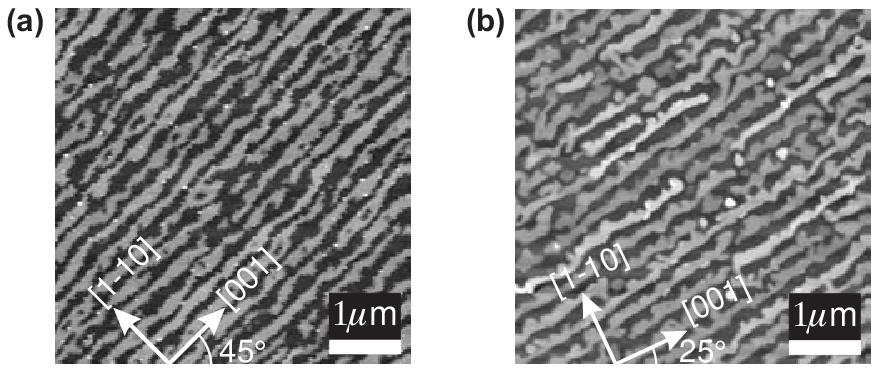}%
\end{center}
  \caption{\label{fig-stripes-with-w}\dataref{From APL 2004}\exsitu AFM images
  of (a)~Fe/W(110) and (b)~Fe/Mo/W(110) stripes obtained by deposition at $\tempC{200}$ followed
  by annealing at $\tempC{450}$.}
\end{figure}

\figref{fig-stripes-with-w} shows Fe/W(110) and Fe/Mo/W(110) nanostructures obtained following the
same procedure as for Mo: deposition at $\tempC{200}$ followed by annealing at \tempC{450}.
Stripes are again formed, however with increased height, centered around $\thicknm{5.5}$. As in
the case of Mo, this suggests that the stripes can be formed thanks to the occurrence of a
metastable height. However the height distribution is larger, and varies slightly with the amount
of material deposited. This suggest that the energy minimum for the preferred height is shallower
for Fe/W(110) than for Fe/Mo(110). It is also evident that the quality of the stripes is higher
for Fe/Mo than for Fe/W. This may stem from the fact that unwetting starts from a film much
thicker for the latter. The quality of the wires could presumably be improved by a detailed
investigation of the shape and depth of grooves formed at the steps during deposition at
$\tempC{200}$, to better assist the unwetting process.

The new process described here, of nanometers-thick stripe formation, is an advancement in
self-organized systems. Indeed the height of metallic stripes and islands produced by mono-atomic
step-decoration was until now limited to one or two
AL\cite{bib-HAU98b,bib-HAU98,bib-SHE97b,bib-VOI91,bib-VOI91b,bib-GAM03b}. In the field of
magnetism this was a fundamental obstacle against the use of such systems for the fabrication of
functional materials, as ferromagnetism was lost at RT because of the low dimensionality. This
obstacle is overcome with the new process\cite{bib-FRU04}. It is therefore of importance to
determine whether this process is specific to Fe(110) or not. As mentioned above stable heights
have been reported for a series of other systems. Besides, grooves at steps or the strong
influence of steps in unwetting processes were reported for other systems like
FeNi/Cu(111)\cite{bib-CHE01} and Cu and Ag/Ru(0001)\cite{bib-LIN04}. We also reported the
influence of atomic steps of the substrate for Tb/Mo(110)\cite{bib-FRU02d}. As these are the two
ingredients that we have identified to yield thick self-organized stripes, it is likely that the
process reported here should occur for other systems.

\subsection{Mono-modal growth of flat islands at very high temperature}

For growth at very high temperature only flat structures form, with no more compact
islands\bracketfigref{fig-wedged-dots}. For Fe/Mo(110) the distribution of heights is again
monomodal, centered around $\thicknm{1.5}$ with however a slightly broader distribution compared
to the growth at lower temperatures. Notice that the top surface of the islands tends to be flat,
yielding wedge-shaped islands. Only when the islands have a very large lateral size steps of mono-
or multi-atomic height are observed on the top surface, which tend to suppress the wedge on long
distances. This again is an indication of the strong stability of the preferred height for
Fe/Mo(110). The statistical examination of cross-sections of the islands reveals that the height
distribution, increased as compared to the case of stripes, primarily arises from the wedge
effect, while the distribution of the \textsl{average} height of the islands is much narrow.
Notice also the tendency of these islands to be elongated along $[1-10]$ with arrow-like ends with
edges along the in-plane $\{111\}$ directions, and that no clear effect of the steps of the
underlying Mo can be evidenced. We believe that the agreement with the height of stripes presented
in the previous sub-section is not fortuitous. However the growth mode at very high temperature is
more complex and needs to be clarified. For example, we cannot exclude that interfacial alloying
between Fe and Mo is involved to some extent in the formation of these islands, as the possibility
of Fe/Mo(110) alloying was reported at high temperature\cite{bib-TIK90}.

Finally, notice that no indication on the type of lateral facets could be obtained from RHEED in
the case of mono-disperse assemblies of thin flat islands, because of the weak area of the facets.
Also, beyond the close completion of a uniform film upon percolation of these islands,
three-dimensional compact dots start to nucleate, with the same aspect ratios as previously
reported.

\subsection{Self-assembled nanobowties}

\begin{figure}
\begin{center}
  \includegraphics[width=144mm]{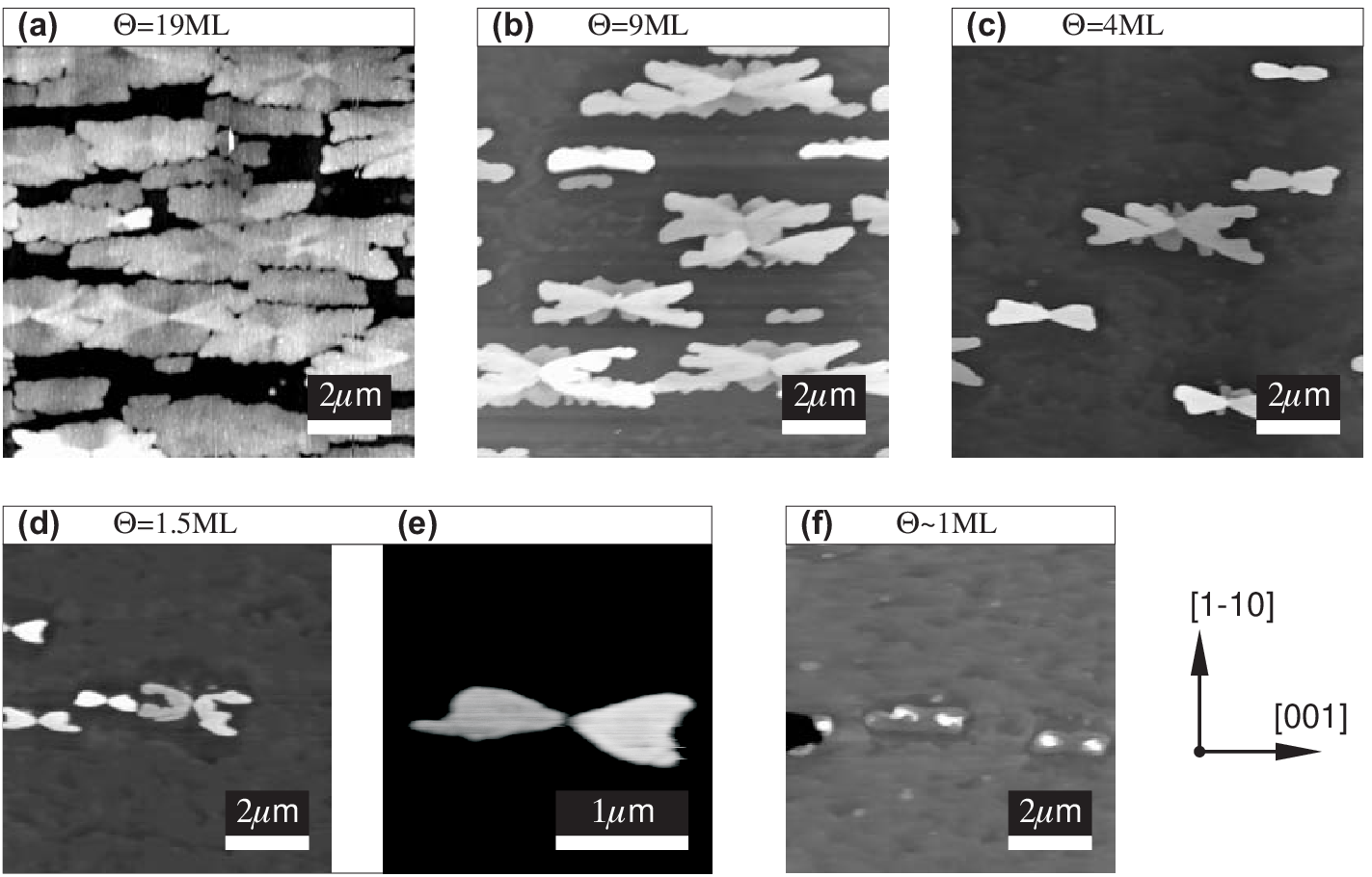}%
\end{center}
  \caption{\label{fig-nanobowties}\dataref{From FRU121, see CDR overview file}\exsitu AFM images
  of Fe grown at \tempC{800} on $\mathrm{W}/\mathrm{Mo}_{10}\mathrm{W}_{90}(110)$.
  Notice the different scale in (e).}
\end{figure}

With a view to understanding the growth process of islands at very high temperature we have
investigated the nucleation regime~(see \figref{fig-nanobowties}. Here the deposition was
performed on a buffer layer of solid solution, see later). The islands then appear as being made
of two main parts, separated by a constriction ever more pronounced for smaller nominal thickness.
The very first stages of nucleation even seem to consist of two disconnected
islands\bracketsubfigref{fig-nanobowties}{f} that progressively merge during subsequent growth.
During the merging process one gets peculiar nanobowtie shapes consisting of two triangular very
flat plateaus of height around a few nanometers, facing each other on either side of a geometrical
constriction. The constriction can be very narrow~(\thicknm{50} or lower) and is two dimensional,
in the sense that the height is also reduced locally. Such objects might prove useful for the
study of magnetic domain walls in constrictions.

\begin{figure}
\begin{center}
  \includegraphics[width=153mm]{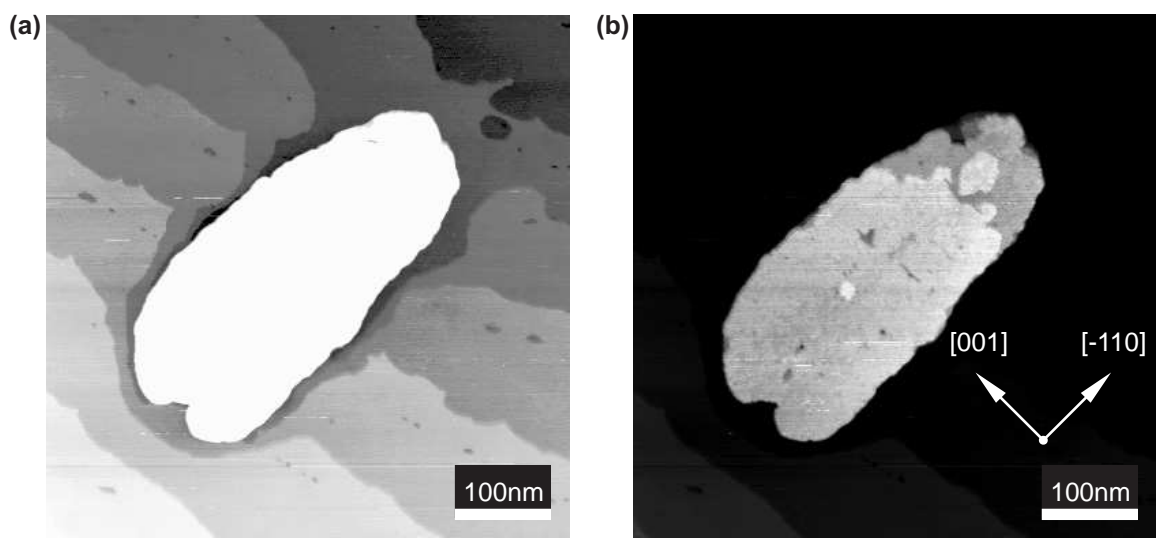}%
\end{center}
  \caption{\label{fig-stm-etching-around-dots}\dataref{Th{\`e}se POJ, Fig.2.8, from chap2/fig-STMIlot800K. PO95/m20}STM image revealing the
  etching process around wedge-shaped flat Fe islands grown on Mo(110) at \tempC{525}. A different color scale is used in (a) and (b), to highlight the
  steps of the buffer layer and the top of the dot, respectively}.
\end{figure}

The surface mobility of metals is so high at these temperatures that the atomistic processes
underlying the nucleation of nanobowties is certainly complex, and remains at the moment unclear.
Notice for instance the depletions occurring around the
nanobowties\bracketfigref{fig-wedged-dots}. The depth of these depletions is larger than one
atomic layer, so that beyond the wetting layer of Fe this suggests material transfer from the
substrate. Whether this material is incorporated in the island or driven away is unknown. Such
erosion effects are known to occur for other materials, provided that they are energetically
favorable and the mobility of atoms is sufficient. Erosion processes are known to occur at room
temperature in the favorable cases like Co/Cu(111)\cite{bib-FIG93,bib-FIG96,bib-PED97} or
Fe/Cu(111)\cite{bib-SHE97b}. As the bare Mo surface already flows step-by-step by 2D evaporation
at the present very high temperatures, it is clear that mobility is sufficient to account for
these effects. Notice that a depletion could already be observed around dots deposited at
intermediate temperature in our experiments\bracketfigref{fig-stm-etching-around-dots} as well as
in the literature~(see \cite{bib-USO05}, Fig.1b). For Fe/Mo(110) the etching of substrate atoms
around the dots is likely to be driven be the strain field arising from Fe, radially compressive
below the dots and extensive at its outer perimeter\cite{bib-MUL05}. Erosion of atoms from the
substrate seems much more limited or even hindered for the case of Fe/W(110), as reported in the
literature\cite{bib-WAC02}.

Finally, it cannot be excluded that minute amounts of contaminants, either of gaseous or solid
origin, may play a role in the nucleation of nanobowties. Indeed the nucleation density is so low
that the level of contamination required to influence the nucleation sites lies well below the
sensitivity of Auger.

\subsection{Physical origin of the preferred height}
\label{subsec-origin-magical-height}

Up to now we have only described the occurrence of preferred heights. In this sub-section we
propose some directions to explain this phenomenon. Let us first reject irrelevant physical
phenomena, before proposing the two most appealing explanations.

First notice that the close coincidence of the height of nanostructures fabricated by deposition
plus annealing on one side, and directly obtained by deposition in the bimodal SK growth mode in a
broad high temperature range or in a monomodal growth mode at very high temperature as reported
previously, strongly suggests that this height stands for a minimum of energy for an Fe(110) film,
and does not result from kinetic effects.

Another argument to reject would be alloying at the interface, with the flat islands being made of
this alloy. This is first in contradiction with the reported complete interface inertness for
Fe/W(110) and very weak reactivity for Fe/Mo(110)\cite{bib-TIK90}. Besides, XRD was performed on
Fe/Mo and Fe/W stripes. Grazing-incidence XRD revealed in both systems the occurrence of the bulk
Fe lattice parameter, with a residual tensile strain lower than $1\%$. Out-of-plane in-lab
characterization could not be carried out on Fe/Mo stripes because the resulting peak is too
broad. For Fe/W the bulk Fe lattice parameter was retrieved with a slight residual tensile strain
of \unit[0.4]{\%}. Most importantly, in this case the out-of-plane crystallographic coherence
length of the nanostructures, as determined with the Scherrer formula from the width of the peak,
yielded $\thicknm{5.5}$ in exact agreement with height inferred from AFM. As a whole this suggests
that the stripes observed by AFM and STM are made of bulk-like Fe in their entire thickness.

Still another argument to reject would be that the bimodal growth might result from size-dependant
strain minimization effects. Indeed cases of bimodal growth are common for semiconductors~(the
so-called domes and huts\cite{bib-MO90,bib-FLO99}) for SK growth modes. This effect was explained
as resulting from the competition of the reduction of surface energy roughly scaling with the
square power of size~(the surface), versus strain energy via \emph{elastic} deformation that
roughly scales with the third power of size~(the volume), thus explaining a transition as a
function of size, which indeed fitted the experimental observations. However this effect requires
a large strain and still coherently-strain islands, which is not the case for Fe/cc(110) flat
islands. Besides, strain is efficiently reduced only in compact islands. Here strain reduction
cannot be a driving force for selecting a value for a uniform height of islands with a lateral
aspect ratio over~100.

The first potential explanation remaining is a quantum size effect, which was recently definitely
proven to give rise to preferred and hindered values in the distribution of
heights\cite{bib-OTE02}. In short, the confinement of electrons along the direction perpendicular
to a layer gives rise to quantum well states~(QWS). The density of energy of the layer is minimum
when the occupied QWS highest in energy is far below the Fermi energy. An argument going in this
direction is the tendency to observe a multimodal distribution of heights both for Fe/Mo(110) and
Fe/W(110), especially in the case of annealing procedures. This can for example be evidenced by
scrutinizing \figref{fig-metastable-thickness} where the flat areas of preferred height are
decorated with a brim of preferred height, however with a larger value. To confirm this we
recently evidenced \insitu with LEEM sudden transitions of height of Fe/W(110)
islands\cite{bib-FRU06b}. If this explanation is correct an open question is why the values of
preferred heights depend so sensitively on the buffer and capping layer, Mo or W, an effect which
could not be explained in the simple model of Ref.\cite{bib-OTE02}.

The second potential explanation is a transition of the network of dislocations at the interface
between Fe and the buffer layer, from a periodic sequence for low thickness to ideally a
non-periodic sequence for large thickness. Indeed in the limit of ultrathin films one expects that
a periodic arrangement of dislocations is the state of lowest energy. This can be understood as
nearest-neighbor interactions dominate in the \textsl{total} energy for low thickness. To the
contrary for very thick films the interface contributes very little to the total energy of the
system. If the dislocation network is periodic a small residual strain remains in the entire
system because the lattice parameters of the two elements are not commensurate. Then, it should be
energetically favorable to reorganize the array of dislocations in a non-periodic sequence at the
cost of a finite interfacial energy, as this allows to lower the strain energy in a majority part
of the system. It is appealing to think that the preferred height could be the critical height for
such a transition. In the case of a commensurate/incommensurate transition one expects a very
rapid variation of behavior as a function of the lattice parameter of the substrate, as what
counts is not directly the misfit, but the residual lateral misfit, \ie the misfit between $N$
atoms of the substrate versus $N+1$ atoms of Fe at the interface. The larger the residual misfit,
the lower the critical thickness would be. Let us give figures. The misfit of Fe with Mo is
$\unit[9.79]{\%}$, as defined by $\epsilon=(a_{\mathrm{Mo}}-a_{\mathrm{Fe}})/a_{\mathrm{Fe}}$, and
that with W is $\unit[10.42]{\%}$. \tabref{tab-residual-misfit} shows the residual misfit expected
for arrays of dislocations with various periods. The table shows that, assuming a positive
residual misfit as is often reported, $N=11$ and $\epsilon=\unit[+0.64]{\%}$ in the case of Fe/Mo,
and $N=10$ and $\epsilon=\unit[+0.38]{\%}$ in the case of Fe/W. These figures are coherent in sign
and magnitude with those determined experimentally by XRD in wires organized along steps. We
mentioned in \secref{sec-compact-dots} that the compacts dots are relaxed to better than
$\unit[0.1]{\%}$, which proves that the arrays of interfacial dislocations evolves at some point.
According to the above figure the elastic energy associated with the residual strain is expected
to be $3-4$ times larger for Fe/Mo than for Fe/W, which would imply the reverse relationship for
the critical thickness, which is the case in the experiments for the preferred height. Despite
this agreement, the possible link with the preferred height remains at this point speculative
until further experimental data is available. Notice that the figures of low strain given above
contrast with surface XRD measurements of the strain for Fe/W(110), which conclude that a positive
strain isotropic in-the-plane remains, equalling \unit[1.22]{\%}\cite{bib-POP03}. The discrepancy
might arise from the fact that these authors studied a \textsl{continuous} film of thickness
\thickAL{12}, grown at \tempK{300} and only slightly annealed. In this case of a continuous film
the array of dislocations if completely formed in the early stages of growth, and the density of
dislocations may not find a pathway to evolve. A different situation is expected for
nanostructures where dislocations can enter or be expelled at the perimeter or nanostructures.
Thus, a comparison of our results with those of Ref.\cite{bib-POP03} may not be done directly.

To decide what explanation is correct between quantum size effects or a commensurate to
incommensurate transition will require calculations and further experiments, both of growth and of
characterization. In the next paragraph we report a route that may help gathering in the future
enough data to solve the matter.

\begin{table}
  \centering
  \caption{Residual misfit of Fe on Mo(110) and W(110) for a periodic array of dislocations, when a series of $N+1$
  Fe atoms coincides with $N$ atoms of the substrate.}
  \begin{tabular}{ccc}
    $N$          &  With Mo &   With W \\\hline
    9  &  \unit[-1.19]{\%}   &  \unit[-0.62]{\%}\\\hline
    10 &  \unit[-0.20]{\%}   &  \unit[+0.38]{\%}\\\hline
    11  &  \unit[+0.64]{\%}   &  \unit[+1.22]{\%}\\\hline
  \end{tabular}
  \label{tab-residual-misfit}
\end{table}

\subsection{Deposition on interfacial layers and solid solutions}

In this paragraph we present preliminary results concerning the growth of Fe nanostructures on the
composite buffer layers described in part~\ref{sec-buffer}.

To prove unambiguously which of the two explanations for the preferred height is correct will
require analyzing the role of several parameters independently, for instance hybridization at the
interface to affect quantum size effects, and lattice misfit to influence the density of
interfacial dislocations. For the latter a continuous variation is desirable because a full curve
provides more information than just a few points.

To vary independently the lattice parameter and the chemical nature of the supporting surface we
apply the procedures reported in section \ref{sec-buffer} for the fabrication of buffer layers.
Concerning the control of the interface we found the following values for critical thickness, as
deduced from self-organized stripes: $t_\mathrm{Fe/W}\sim\thicknm{5.5}$,
$t_\mathrm{Fe/Mo}\sim\thicknm{1.4}$, $t_\mathrm{Fe/Mo/W}\sim\thicknm{6}$,
$t_\mathrm{Fe/W/Mo}\sim\thicknm{4.5}$. It is already clear that both the in-plane lattice
parameter and the chemical nature of the interface, influence the metastable thickness of the
stripes.

\begin{figure}
\begin{center}
  \includegraphics[width=160mm]{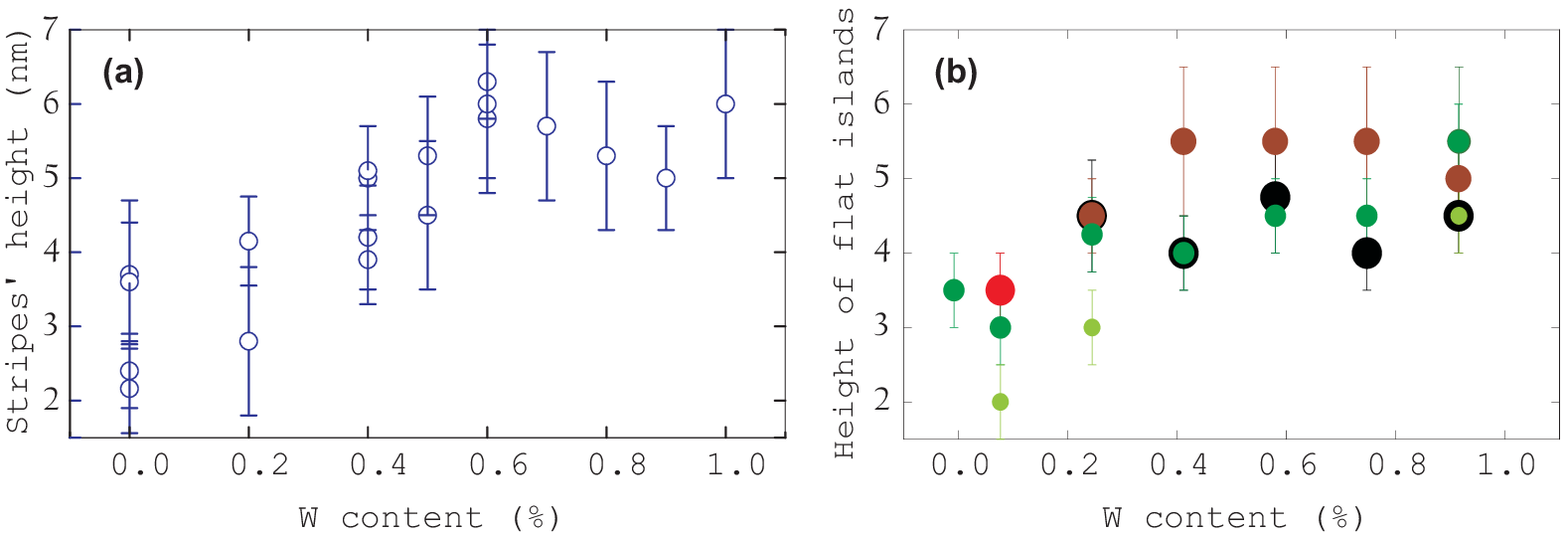}%
\end{center}
  \caption{\label{fig-preferred-height-with-ss}\dataref{(a)~Extracted from Vortrag-Hamburg 2005 (and before?)
  (b)~AFM/Fru121.OPJ} Height of Fe(110) nanostructures obtained upon deposition on $\mathrm{Mo}(\thickAL{3})/\mathrm{W}_x\mathrm{Mo}_{1-x}$
  composite buffer layers. For both curves the error bars stand for the full width of the experimental distribution over the assembly
  of nanostructures. (a)~Stripes decorating steps, upon deposition at $\tempC{200}$ and annealing at $\tempC{500}$
  (b)~flat islands upon deposition at $\tempC{800}$. For the latter case the distribution of height is multimodal, the height being
  larger for the area related to the initial stages of growth with a shape of bow tie\bracketfigref{fig-nanobowties}.
  In the present figure we plot the height of the remaining part of these dots, which is flat, monomodal and does not vary significantly
  upon deposition. The diameter of the symbols stands for the nominal thickness $\Theta$ of Fe, ranging from zero to \thicknm{4}.
  The color, constant for each $\Theta$, is used for clarity when several symbols overlap.}
\end{figure}

Concerning the variation of lattice parameter \tabref{tab-residual-misfit} shows that there exists
a concentration $x$ of the solid solution $\mathrm{W}_x\mathrm{Mo}_{1-x}$ for which the misfit
cancels out. For this concentration in principle the critical thickness should go to infinity for
commensurate lattices, although in a real experiment other effects will interfere. The
differential thermal expansion between Sapphire and the metals should also be taken in to account.
In \figref{fig-preferred-height-with-ss} we plot the preferred height of nanostructures as a
continuous function of the lattice parameter, for the interface $\mathrm{Mo}(\thickAL{3})$.
\figref{fig-preferred-height-with-ss}a displays figures concerning stripes obtained upon
deposition at moderate temperature followed by annealing. \figref{fig-preferred-height-with-ss}b
displays values concerning islands obtained upon deposition at very high temperature. First notice
the remarkable agreement between the two curves. This suggests that the preferred height really
comes from an energetic minimum rather that results from a kinetic effect, as the two fabrication
procedures are very different. Besides, the height varies significantly with the lattice parameter
of the buffer layer. However no clear maximum is observed, which would have been the experimental
signature for a divergence expected theoretically. At this point our experiments are therefore not
conclusive concerning the role of lattice parameter. Extension of these results to a variation of
lattice parameter much beyond the range offered by mixtures of Mo and W are scheduled.

\section{Discussion of growth processes with respect to existing data}

In this section we briefly discuss our results with respect to the literature. We concentrate on
new aspects and/or differences.

Concerning the stripes we could find no reports of annealing of Fe/Mo(110) films prior to our
experiments. However several reports existed for annealed Fe/W(110) films, so that the question
arises why self-organization of stripes had not been evidenced. In fact some wires were reported
for Fe/W(110)\cite{bib-SAN99,bib-ROU06}. However these wires differ from ours by two aspects.

The first differing aspect is that the wires reported in the literature are always roughly
oriented along $[001]$, whatever the step orientation. The [001] orientation upon annealing is
likely to result from anisotropic surface diffusion\cite{bib-REU98,bib-KOH00} during the unwetting
process. The reason why the atomic steps of the substrate were not effective is certainly that
growth was performed at RT. In these conditions we have shown that grooves are not formed, so that
the memory of the steps of the buffer surface is lost. In the case of Mo we evidenced a
preferential alignement of nanostructures forming upon annealing films deposited at RT, for
\thickAL{2-3}\bracketsubfigref{fig-metastable-thickness}{a-b}. This probably happens because this
thickness is still comparable to the height of a step. However for W the steps can play no role
because the preferred height is much larger than for Mo, so that at the minimum height required
for the formation of stripes~($\thicknm{2}$, not illustrated with pictures here) the influence of
buried steps is negligible. In recent experiments monitored \insitu with LEEM we have shown that
the formation of stripes is however easier along steps roughly parallel to $[001]$, probably
resulting from the anisotropic diffusion afore mentioned\cite{bib-FRU06b}.

The second differing aspect if that the height of the reported stripes was generally larger than
$\thicknm{5.5}$. The height, width and spacing of the wires also varied with $\Theta$ and
annealing conditions. This is typical for an annealing process without a preferred height.
Incidentally, we sometimes observe such thicker wires amid the self-organized array. They seem to
have grown at the expense of neighboring wires, and are therefore much thicker. The occurrence of
such wires much depends on kinetics during the annealing.

Concerning islands there are several articles that report on them, both for W and Mo buffer
layers. However as these were reported in the first stages of growth~(with respect to our
results), it is not straightforward to say to what type they belong, or even whether such a
classification makes sense at all for small islands. The discussion should therefore be taken with
care. First, all authors insist that their islands are wedge-shaped. Our compact islands are also
wedge-shaped because their top is nearly atomically flat, however the wedge feature is in relative
value not striking owing to the large thickness of the dots. Thus, the fact to be wedged does not
provide a mean for discrimination. Let us now examine figures for the aspect ratio. The first
report of wedge-shaped islands for Fe/W(110) is by Bethge \etal\cite{bib-BET95}, upon growth at
\tempC{300}. The average thickness of the islands remains around \thicknm{1-1.5} so the comparison
with both compact islands and the preferred height is difficult. $\eta\sim0.01$ and $r\sim1$ can
be read from the published images. Wachowiak \etal also fabricated islands on W by depositing Fe
at RT, followed by annealing at $\tempC{525\pm100}$\cite{bib-WAC02}. The aspect ratios of the
island studied are $\eta\sim0.037$ and $r\sim1.93$, \ie compatible with the ranges of ratios found
for our compact islands on~W\bracketfigref{fig-plot-aspect-ratios}. Again on W, R{\"o}hlsberger \etal
fabricated islands by depositing Fe at RT, followed by annealing at $\tempC{425}$\cite{bib-ROH03}.
Quoting the authors the height of the islands is $h=\thicknm{3.9\pm0.8}$. The vertical aspect
ratio is $r\sim0.019$, clearly out of the range from \figref{fig-plot-aspect-ratios}, and the
length is pretty much distributed, with $\eta$ ranging from 1 to 5. The characteristics are thus
closer to 'flat' islands, although the thickness is slightly smaller than what we report in this
manuscript. Still on W, Sander \etal obtained thick wires aligned along $[001]$. Quoting these
authors, these \emph{were observed either after growth at \tempK{1000}, or after annealing at
\tempK{700-1000} a room-temperature-grown Fe film}. Thus, there seems to be a contradiction with
R{\"o}hlsberger for annealing around $\tempC{425}$, unless the temperature calibration is different.
Besides, we have never obtained wires by deposition at high temperature, so that this point is
also unclear. Concerning Mo Malzbender \etal fabricated islands by deposition at $\tempC{325}$.
The islands have a \emph{maximum height of \thickAL{10}} which is very comparable to the report of
Bethge \etal quoted above, and compatible with the preferred height we observe. Again on Mo,
Murphy \etal came to similar results for deposition at $\tempC{220}$. However their results may
not be directly comparable since their substrate is more vicinal, with a terrace size around
\thicknm{20}.

To conclude the discussion, islands have been reported in the literature for both Fe/Mo and Fe/W,
but at too early a stage to be unambiguously classified as 'compact' or 'flat'. Besides, some
contradictory results are found between different reports in the literature, or between literature
and our work. Therefore work is still needed to clarify the situation. When making comparison, one
should also keep in mind that the characteristics of flat wedge-shaped islands (density, size,
shape and orientation) are certainly influenced by the density and orientation of the atomic steps
of the substrate, as it was clearly shown that these islands nucleate from the
steps\cite{bib-MUR02}. The examination of published data indeed shows that the density of dots
increases with the density of atomic steps, and so at the same time the height of the dots for a
given growth procedure decreases. An increased density of the nucleation of dots is a common
feature, see \eg \cite{bib-OTE02}. Also, uncontrollable minute amounts of contaminants may play a
role in the nucleation of stripes versus dots.

Finally, no data is available in the literature concerning the growth at very high temperature.
The question why no compact islands grow from the beginning in this case is still open. One reason
might be related to the high mobility of adatoms, causing the meeting of adatoms on the top of
flat islands to be improbable. Thus, the nanostructures could grow only laterally by aggregation
of adatoms and movement of kinks and steps. In the case of intermediate temperature, the
non-hindered nucleation would be explained by the significant probability of adatoms to meet on
top of the flat nanostructures and form a stable nucleus. It was also shown that adatom dynamics
could vary by orders of magnitude in relation to strain\cite{bib-TSI03}. It is therefore an open
possibility that strain and stress at the edge of islands, which is known to be very high, plays a
role.

\section{Magnetic investigations}

The growth and structural aspects of our investigations of Fe(110) self-assembled nanostructures,
which were detailed above, had only been reported partly in previous papers. In contrast most of
our studies of magnetic properties of Fe(110) have already been
published\cite{bib-FRU02c,bib-FRU03c,bib-FRU04,bib-FRU04c,bib-FRU05}. Therefore we propose here
only a condensed overview of our investigations. Details can be found in the original papers.

As a natural route to magnetic nanostructures we first investigated magnetic
films\cite{bib-FRU99c}. The magnetic anisotropy of Fe/W(110) films with various capping layers had
been extensively
studied\cite{bib-GRA86,bib-HIL87,bib-ELM90,bib-BAU91,bib-ELM91,bib-GRA93,bib-GER93,bib-FRI94,bib-ELM94b}.
To the contrary, nearly no data was available for UHV-grown Fe/Mo(110) films nor W/Fe/W and
Mo/Fe/Mo trilayers before our studies\cite{bib-FRU99c}, apart from a report on Fe/Mo(110)
multilayers, besides with significant uncertainties about surface anisotropy
values\cite{bib-OSG95b}. Both W/Fe/W and Mo/Fe/Mo films keep their magnetization in-the-plane for
any thickness. Thick Fe(110) films, typically several tens of nanometers, display a uniaxial
anisotropy in-the-plane with $[001]$ as an easy axis resulting from the projection of the bulk
cubic anisotropy in the $(110)$ plane. At lower thicknesses the relative contribution of
interface-like anisotropy increases. For W/Fe/W films the interfacial contribution favors the
$[1-10]$ axis, resulting in a reorientation transition around $\thicknm{10}$, similarly to
UHV/Fe/W(110) films\cite{bib-GRA93,bib-GER93}. For Mo/Fe/Mo films the interfacial contribution
favors the $[100]$ axis. This strengthens the bulk anisotropy and thus does not give rise to a
reorientation transition. Notice that this contrasts with uncapped structures, where
Fe(\thickAL{1}) on calculations predict an easy axis along $[1-10]$ on both
W(110)\cite{bib-GAL00,bib-QIA01} and Mo(110)\cite{bib-QIA01}, while only the former is document
experimentally. Below $\thicknm{2-3}$ the anisotropy becomes dominantly of second order with an
anisotropy field around $\unit[0.5]{T}$ for $\thicknm{1}$-thick films. We have checked that cc/Fe
and Fe/cc interfaces induce a qualitatively similar change of interfacial anisotropy, \ie of same
sign and similar value\cite{bib-FRUunpub}. Thus W/Fe/Mo and Mo/Fe/W films display a moderate total
interfacial anisotropy, dominated by that of W. Thus, in principle using mixtures of Mo and W for
buffer or capping layers should allow one to tune the effective anisotropy of Fe(110) layers
independently from their thickness.

The first series of magnetic studies is concerned with thick stripes. Our interest in thick
stripes arose from the fact that all self-organized systems reported so far had a height of one or
two atomic layers\cite{bib-HAU98b,bib-HAU98,bib-SHE97b,bib-VOI91,bib-VOI91b,bib-GAM03b}. For such
small dimensions thermal excitations become dominant over anisotropy, and no magnetic properties
such as coercivity and remanence had been demonstrated at room temperature. A handwavy
argument\cite{bib-FRU05b} shows that the effect of thermal excitations increases faster than
interfacial anisotropy, so that the effect becomes ever more acute for smaller dimensions, as is
reported\cite{bib-GAM02,bib-GAM03b}. This argument remains valid whatever the type of interface
be, surfaces, edges or kinks depending on the dimensionality. We issued a first report on
irregular and discontinuous Mo/Fe/Mo stripes, with magnetic functionality at room
temperature\cite{bib-FRU02c}. We improved the growth process of stripes, as reported above, which
resulted in the fabrication of continuous stripes of height in the range $\thicknm{3-5}$ with
various buffer layers and cappings. For stripes aligned along $[001]$, and buffer and capping
interfaces made of Mo, both dipolar and interface anisotropy add up. This resulted in a coercivity
up to $\unit[50]{mT}$ and a high remanence along the $[001]$ direction. Stable magnetic domains of
size a few hundreds of nanometers can be formed upon demagnetization along in-plane
$[1-10]$\cite{bib-FRU04}.

The second series of magnetic studies is concerned with compact dots, mostly Mo/Fe/Mo. In the case
of W/Fe/W in the range of thickness $[\thicknm{5-20}]$ the anisotropy is rather low and with a
symmetry close to fourfold, owing to the occurrence of a reorientation transition, see the above
paragraph dealing with the anisotropy of continuous films. Thus such dots are magnetically soft
and display various types of flux-closure states\cite{bib-WAC02,bib-BOD04b} expect in the
ultrathin range\cite{bib-RAM98}. The magnetic anisotropy is essentially uniaxial and stronger in
the case of Mo/Fe/Mo in this range of thickness, so that the dots are mostly in a single domain
state and display a significant remanence\cite{bib-FRU01b}. For a larger thickness Mo/Fe/Mo dots
display flux-closure states. In the approximate range $[\thicknm{30-80}]$ both Landau~(one single
Bloch wall) and diamond~(two or more vortices) states occur\cite{bib-FRU03c}. The shape of the
domains is well reproduced by the two-dimensional model of Van den Berg\cite{bib-VAN84,bib-VAN86}.
The observation of either a Landau or a diamond state in dots of similar shapes could be explained
by the role of an asymmetry, like that of shape, during the demagnetization process: in a
symmetric dot two vortices are created on opposite locii ending up in a diamond state, while only
one vortex ending up in a Landau state may enter a significantly asymmetric dot\cite{bib-FRU04c}.
For thicknesses in the range $[\thicknm{80-250}]$ only Landau states are found, which can be
explained by the increasing cost of N{\'e}el walls, which are found in a diamond state. Besides, as
the thickness of such dots is now comparable or larger than domain walls, the magnetization state
varies along the thickness of the dot, and deviate from the model of Van den Berg. Magnetization
flux closure can be evidenced both in-the-plane and along vertical cross-sections: these dots,
although of size in the micrometer range, are premisses of bulk flux-closure states. This leads to
unusual features like asymmetric N{\'e}el walls. These are forbidden in continuous films for such
large thicknesses, but are stabilized in dots owing to the geometrical constraint. Such dots are
therefore a model system to bridge the gap between simple nanostructure now well understood like
thin flat dots, and bulk materials.

\section*{Conclusion}

We have revisited the growth of Fe on W(110) and Mo(110) surfaces, to which this article is
devoted. The first new key feature is that growth, which was already known to be of
Stranski-Krastanov type, is here shown to be characterized by a bimodal distribution of islands.
One species consists of compact facetted islands that grow homothetically during deposition, and
whose shape can be understood with the Wulff-Kaischev construction. The interface energy between
Fe and the substrate can be extracted by analyzing the shape of the dots, which yields
$\gamma_\mathrm{int,Mo}=\unit[0.65\pm0.15]{J.m^{-2}}$ and
$\gamma_\mathrm{int,W}=\unit[0.10\pm0.05]{J.m^{-2}}$. Notice that these are effective values that
include the strain energy located close to the interface, arising from the core and surroundings
of interfacial misfit dislocations. The second species consists of thin flat islands with a
preferred height: $\thicknm{1.4\pm0.2}$ for Fe/Mo and a broader distribution centered around
$\thicknm{5.5}$ for Fe/W. Using W-Mo solid solutions we fabricated buffer layers whose lattice
parameter can be continuously adjusted, and whose surface can be either pure W or pure Mo by using
an ultrathin pseudomorphic capping layer. The preferred height is influenced by both parameters.
Two possible origins for the preferred height are quantum-size effects and
commensurate/discommensurate transition of the interfacial dislocations network. Further works is
needed to determine which effect is relevant. Finally, we took advantage of the preferred height
to fabricate arrays of nanometers-thick stripes self-organized along the atomic steps of the
buffer layer. The orientation and separation of the stripes can be adjusted by controlling the
miscut angle and azimuth, their width by the amount of material deposited, and their height can be
controlled via slight variations of the lattice parameter of solid-solution buffer layers.

As an outlook we believe that the principle of growing buffer layers of solid solutions
$\mathrm{A}_x\mathrm{B}_{1-x}$ with a full range of $x$ from 0 to 1, which we named
\textsl{chemical wedge} or \textsl{chemical gradient layer}~(CGL), is a promising tool. This
process is inspired from the concept of wedges of \textsl{thickness} that is nowadays used by many
groups, which allows one to perform quick and systematic studies of the influence of thickness of
some feature, be it of growth or of a physical property. Chemical wedged should allow the study of
the effects related to misfit, either for growth or physical properties.

\section*{Appendix~I}
\addcontentsline{toc}{section}{Appendix~I}%

The Wulff-Kaischev construction relates the occurrence and geometry of the facets arising in a
supported crystal, with the values of surface and interface energies~[\eqnref{eq-WulffKaichev}].
Based on this construction, straightforward calculations yield the vertical and lateral aspect
ratios expected for bcc(110) supported crystal. If only $\{110\}$ and $\{001\}$ facets are
considered one finds:

{
\def\Ea{\gamma_{\{110\}}}
\def\Eb{\gamma_{\{001\}}}
\begin{itemize}
  \item For $0<\gammaint<\Ea(\sqrt{2}-1)$ ($\simeq\unit[1]{J/m^{2}}$)
  \begin{eqnarray}
    r=\frac{L}{l} &=& \frac{1}{\sqrt{2}}\frac{(\gammaint+\Ea)}{(\gammaint-\Ea+\sqrt{2}\Eb)}\\
    \eta=\frac{h}{l} &=& \frac{1}{2}\frac{\gammaint}{(\gammaint-\Ea+\sqrt{2}\Eb)}
  \end{eqnarray}

  \item For $\Ea(\sqrt{2}-1)<\gammaint<2\Ea-\sqrt{2}\Eb$ ($\simeq\unit[1.72]{J/m^{2}}$)
  \begin{eqnarray}
    r=\frac{L}{l} & = & \frac{\Ea}{(\gammaint-\Ea+\sqrt{2}\Eb)}\\
    \eta=\frac{h}{l} & = & \frac{1}{2}\frac{\gammaint}{(\gammaint-\Ea+\sqrt{2}\Eb)}
  \end{eqnarray}

  \item For $2\Ea-\sqrt{2}\Eb<\gammaint<\Ea$ ($=\unit[2.43]{J/m^{2}}$)
  \begin{eqnarray}
    r=\frac{L}{l} & = & 1\\
    \eta=\frac{h}{l} & = & \frac{\gammaint}{2\Ea}.
  \end{eqnarray}

\end{itemize}
}

More general formulae taking into account $\{310\}$ and $\{211\}$ facets are given in
Ref.\cite{bib-JUB01b}. In the case of Fe(110) these change only slightly the value of $r$, while
$\eta$ remains nearly unchanged. The numerical values\bracketfigref{fig-plot-aspect-ratios} are
derived using the surface energies computed in Ref\cite{bib-VIT98}.

\section*{Appendix~II}
\addcontentsline{toc}{section}{Appendix~II}%

In this appendix we calculate the angles $\beta$ (with respect to the substrate plane) of the
possible facets of bcc(110) dots for various azimuths of the electron beam. Let us write
$\vect{a}=a\vect{i}$, $\vect{b}=a\vect{j}$ and $\vect{c}=a\vect{k}$ the three vectors of the cubic
cell in real space, $\vect{a^*}=(2\pi/a)\vect{i}$, $\vect{b^*}=(2\pi/a)\vect{j}$ and
$\vect{c^*}=(2\pi/a)\vect{k}$ the reciprocal lattice vectors. Let us write a scattering vector in
the general form: $\vect{q}=h\vect{a^*}+k\vect{b^*}+l\vect{c^*}$. Only the nodes for which
$h+k+l\equiv0 [2]$ arise for a simple centered cubic cell.  For a given azimuth $\vect R$ RHEED
patterns may arise only from facets on which the electron beam impinges at grazing incidence, that
is for $\vect{q}.\vect{R}=0$. The cases of three azimuths are treated below, for which facets
indeed arise for our Fe(110) dots.

\subsection{Azimuth $[001]$}

Let us look for planes determined by $\vect{q}.\vect{c}=0$, thus satisfying $l=0$. The general
index of these is thus $(hk0)$ with $h+k\equiv0 [2]$. One has $\cos\beta=(h+k)/\sqrt{2(h^2+k^2)}$,
$\sin\beta=(h-k)/\sqrt{2(h^2+k^2)}$ and $\tan\beta=(h-k/h+k)$. The features of the main facets are
reported in \tabref{tab-azimuthZZO}.

\begin{table}[h]
  \centering
  \caption{Angles of the main possible facets for the electron beam azimuth $[001]$.}
  \begin{tabular}{ccc}\hline\hline
    $(hk)$  &  Plane $(hkl)$ &   Angle with surface \\\hline
    $(24)$  &  $(240)$   &  $\angledeg{\sim18.43}$ \\
    $(13)$  &  $(130)$   &  $\angledeg{\sim26.57}$ \\
    $(02)$  &  $(020)$   &  $\angledeg{45}$ \\\hline\hline
  \end{tabular}
  \label{tab-azimuthZZO}
\end{table}

\subsection{Azimuth $[1-10]$}

Let us look for planes determined by $\vect{q}.(\vect{a-b})=0$, thus satisfying $h=k$. The general
index of these is thus $(hhl)$ with $2h+l\equiv0 [2]$. One has $\cos\beta=2h/\sqrt{2(2h^2+l^2)}$,
$\sin\beta=l/\sqrt{2h^2+l^2}$ and $\tan\beta=l/(h\sqrt{2})$. The features of the main facets are
reported in \tabref{tab-azimuthOOZ}.

\begin{table}[h]
  \centering
  \caption{Angles of the main possible facets for the electron beam azimuth $[1-10]$.}
  \begin{tabular}{ccc}\hline\hline
    $(hl)$  &  Plane $(hkl)$ &   Angle with surface \\\hline
    $(3\overline{2})$  &  $(33\overline{2})$   &  $\angledeg{\sim25.23}$ \\
    $(2\overline{2})$  &  $(22\overline{2})$   &  $\angledeg{\sim35.26}$ \\
    $(1\overline{2})$  &  $(11\overline{2})$   &  $\angledeg{\sim54.74}$ \\\hline\hline
  \end{tabular}
  \label{tab-azimuthOOZ}
\end{table}

\subsection{Azimuth $[1-1-1]$}

Let us look for planes determined by $\vect{q}.(\vect{a-b-c})=0$, thus satisfying $h-k=l$. The
general index of these is thus $(h,k,h-k)$ with $2h\equiv0 [2]$, so with no restriction on~$h$.
One has $\cos\beta=(h+k)/[2\sqrt{(h^2+k^2-hk)}]$, $\sin\beta=(\sqrt{3}/2)(h-k)/\sqrt{h^2+k^2-hk}$
and $\tan\beta=(\sqrt{3})(h-k/h+k)$. The features of the main facets are reported in
 \tabref{tab-azimuthOOO}.

\begin{table}[h]
  \centering
  \caption{Angles of the main possible facets for the electron beam azimuth $[1-1-1]$.}
  \begin{tabular}{ccc}\hline\hline
    $(hk)$  &  Plane $(hkl)$ &   Angle with surface \\\hline
    $(23)$  &  $(23\overline{1})$   &  $\angledeg{\sim19.11}$ \\
    $(12)$  &  $(12\overline{1})$   &  $\angledeg{30}$ \\
    $(13)$  &  $(13\overline{2})$   &  $\angledeg{\sim40.89}$ \\
    $(01)$  &  $(01\overline{1})$   &  $\angledeg{60}$ \\\hline\hline
  \end{tabular}
  \label{tab-azimuthOOO}
\end{table}

\section*{Acknowledgments}

We are happy to acknowledge technical help concerning electronics~(S. Biston, S. Pelle), XRD~(L.
Ortega) wafer and target preparation~(D. Barral, K. Ayadi, R. Haettel). Concerning magnetism, we
acknowledge the contribution of co-authors in the articles cited in the present condensed
presentation: J.~C. Toussaint, B. Kevorkian, Y. Samson, J. Vogel, A. Locatelli, A. Ballestrazzi,
W. Wernsdorfer, R. Hertel, J. Kirschner, D. Mailly, S. Cherifi, S. Heun. We are finally grateful
to B. Canals and M. Taillefumier for help with Mathematica.

This work was partly funded by \textsl{R{\'e}gion Rh{\^o}ne-Alpes} (Emergence 2001) and the French
ministry of Research (ACI Nanostructures 2000).

\section*{References}


\begin{thebibliography}{100}

\bibitem{bib-GRA82}
U. Gradmann and G. Waller, \ss {\bf 116},  539  (1982).

\bibitem{bib-TIK90}
M. Tikhov and E. Bauer, \ss {\bf 232},  73  (1990).

\bibitem{bib-VIT98}
L. Vitos, A.~V. Ruban, H.~L. Skiver, and J. Koll\'ar, \ss {\bf 411},  186
  (1998).

\bibitem{bib-MAL98}
J. Malzbender, M. Przybylski, J. Giergiel, and J. Kirschner, \ss {\bf 414},
  187  (1998).

\bibitem{bib-KOL00}
J. Kolaczkiewicz and E. Bauer, \ss {\bf 450},  106  (2000).

\bibitem{bib-REU98}
D. Reuter, G. Gerth, and J. Kirschner, \prb {\bf 57},  2520  (1998).

\bibitem{bib-SLA04}
M. Sladecek, B. Sepiol, J. Korecki, T. Slezak, R. R{\"u}ffer, D. Kmiec, and G.
  Vogl, \ss {\bf 566-568},  372  (2004).

\bibitem{bib-FRU03d}
P.~O. Jubert, O. Fruchart, and C. Meyer, \ss {\bf 522},  8  (2003).

\bibitem{bib-GAR83}
T.~M. Gardiner, \tsf {\bf 105},  213  (1983).

\bibitem{bib-ELM94}
H.~J. Elmers, J. Hauschild, H. H{\"o}che, U. Gradmann, H. Bethge, D. Heuer, and
  U. K{\"o}hler, \prl {\bf 73},  989  (1994).

\bibitem{bib-BET95}
H. Bethge, D. Heuer, C. Jensen, K. Resh{\"o}ft, and U. K{\"o}hler, \ss {\bf
  331-333},  878  (1995).

\bibitem{bib-SAN99b}
D. Sander, A. Enders, and J. Kirschner, \epl {\bf 45},  208  (1999).

\bibitem{bib-MEY01}
H.~L. Meyerheim, D. Sander, R. Popescu, and J. Kirschner, \prb {\bf 64},
  045414  (2001).

\bibitem{bib-BOD96}
M. Bode, R. Pascal, M. Dreyer, and R. Wiesendanger, \prb {\bf 54},  R8385
  (1996).

\bibitem{bib-POP03}
R. Popescu, H.~L. Meyerheim, D. Sander, J. Kirschner, P. Steadman, O. Robach,
  and S. Ferrer, \prb {\bf 68},  155421  (2003).

\bibitem{bib-MUR02}
S. Murphy, D. {Mac Math\'una}, G. Mariotto, and I.~V. Shvets, \prb {\bf 66},
  195417  (2002).

\bibitem{bib-MAE04}
Y. Maehara, A. Yamada, H. Kawanowa, and Y. Gotoh, \ass {\bf 237},  316  (2004).

\bibitem{bib-CLE93}
B.~M. Clemens, R. Osgood, A.~P. Payne, B.~M. Lairson, S. Brennan, R.~L. White,
  and W.~D. Nix, \jmmm {\bf 121},  37  (1993).

\bibitem{bib-FRU99c}
O. Fruchart, J.-P. Nozi\`eres, and D. Givord, \jmmm {\bf 207},  158  (1999).

\bibitem{bib-ALB93}
M. Albrecht, H. Fritzsche, and U. Gradmann, \ss {\bf 294},  1  (1993).

\bibitem{bib-FRU98b}
O. Fruchart, S. Jaren, and J. Rothman, \ass {\bf 135},  218  (1998).

\bibitem{bib-KOH00}
U. K{\"o}hler, C. Jensen, C. Wolf, A.~C. Schindler, L. Brendel, and D. Wolf,
  \ss {\bf 454-456},  676  (2000).

\bibitem{bib-HAU98}
J. Hauschild, U. Gradmann, and H.~J. Elmers, \apl {\bf 72},  3211  (1998).

\bibitem{bib-HAU98b}
J. Hauschild, H.~J. Elmers, and U. Gradmann, \prb {\bf 57},  R677  (1998).

\bibitem{bib-ELM00}
H. Elmers, J. Hauschild, and U. Gradmann, \jmmm {\bf 221},  219  (2000).

\bibitem{bib-BER90}
P.~J. Berlowitz, J.~W. He, and D.~W. Goodman, \ss {\bf 231},  315  (1990).

\bibitem{bib-SAN98}
D. Sander, R. Slomski, A. Enders, C. Schmidthals, D. Reuter, and J. Kirschner,
  \jpd {\bf 31},  663  (1998).

\bibitem{bib-SAN98b}
D. Sander, A. Enders, C. Schmidthals, D. Reuter, and J. Kirschner, \ss {\bf
  402-404},  351  (1998).

\bibitem{bib-SAN99}
D. Sander, \rpp {\bf 62},  809  (1999).

\bibitem{bib-WAC02}
A. Wachowiak, J. Wiebe, M. Bode, O. Pietzsch, M. Morgenstern, and R.
  Wiesendanger, \science {\bf 298},  577  (2002).

\bibitem{bib-ROH03}
R. R{\"o}hlsberger, J. Bansmann, V. Senz, K.~L. Jonas, A. Bettac, and K.~H.
  Meiwes-Broer, \prb {\bf 67},  245412  (2003).

\bibitem{bib-USO05}
V. Usov, S. Murphy, and I. Shvets, \jmmm {\bf 286},  18  (2005).

\bibitem{bib-FRU05b}
O. Fruchart, \crp {\bf 6},  61  (2005).

\bibitem{bib-SHE04}
J. Shen, {Zheng Gaib}, and J. Kirschner, \ssr {\bf 52},  163  (2004).

\bibitem{bib-FRU01b}
P.-O. Jubert, O. Fruchart, and C. Meyer, \prb {\bf 64},  115419  (2001).

\bibitem{bib-FRU01c}
P.~O. Jubert, O. Fruchart, and C. Meyer, \jmmm {\bf 226-230},  1842  (2001).

\bibitem{bib-FRU02b}
O. Fruchart, P.-O. Jubert, C. Meyer, M. Klaua, J. Barthel, and J. Kirschner,
  \jmmm {\bf 239},  224  (2002).

\bibitem{bib-FRU02c}
P.-O. Jubert, O. Fruchart, and C. Meyer, \jmmm {\bf 242-245},  565  (2002).

\bibitem{bib-FRU04}
O. Fruchart, M. Eleoui, J. Vogel, P.-O. Jubert, A. Locatelli, and A.
  Ballestrazzi, \apl {\bf 84},  1335  (2004).

\bibitem{bib-CHE93b}
N. Cherief, D. Givord, O. McGrath, Y. Otani, and F. Robaut, \jmmm {\bf 126},
  225  (1993).

\bibitem{bib-TYR94}
G.~C. Tyrrell, T. York, N. Cherief, D. Givord, J. Lawler, J.~G. Lunney, M.
  Buckley, and I.~W. Boyd, \me {\bf 25},  247  (1994).

\bibitem{bib-FRU02d}
S. Pokrant, O. Fruchart, C. Meyer, and L. Ortega, \ss {\bf 506},  235  (2002).

\bibitem{bib-NIS86}
Y. Nishihata, M. Nakayama, H. Kato, N. Sano, and H. Terauchi, \jap {\bf 60},
  3523  (1986).

\bibitem{bib-RUS99}
S. Rusponi, G. Costantini, F. {Buatier de Mongeot}, C. Boragno, and U. Valbusa,
  \apl {\bf 75},  3318  (1999).

\bibitem{bib-VAL02}
U. Valbusa, C. Boragno, and F. {Buatier de Mongeot}, \jpcm {\bf 14},  8153
  (2002).

\bibitem{bib-SEK03}
D. Sekiba, R. Moroni, G. Gonella, F. {Buatier de Mongeot}, C. Boragno, L.
  Mattera, and U. Valbusa, \apl {\bf 84},  762  (2003).

\bibitem{bib-SWI99}
W. \'Swie\c{c}h, M. Mundschau, and C. Flynn, \ss {\bf 437},  61  (1999).

\bibitem{bib-MAY01}
U. May, R. Calarco, J.~O. Hauch, H. Kittur, M. Fonine, U. R{\"u}diger, and G.
  G{\"u}ntherodt, \ss {\bf 489},  144  (2001).

\bibitem{bib-FON03}
M. Fonin, Y.~S. Dedkov, J. Mayer, U. R{\"u}diger, and G. G{\"u}ntherodt, \prb
  {\bf 68},  045414  (2003).

\bibitem{bib-VAL05}
D.~A. Valdaitsev, A. Kukunin, J. Prokop, H.~J. Elmers, and G. Sch{\"o}nhense,
  \apa {\bf 80},  731  (2005).

\bibitem{bib-FRA06}
M. Fraune, J.~O. Hauch, G. G{\"u}ntherodt, M. Laufenberg, M. Fonin, U.
  R{\"u}diger, J. Mayer, and P. Turban, \jap {\bf 99},  033904  (2006).

\bibitem{bib-TUR02b}
P. Turban, S. Andrieu, B. Kierren, E. Snoeck, C. Teodorescu, and A. Traverse,
  \prb {\bf 65},  134417  (2002).

\bibitem{bib-FRI00}
P.~Y. Friot, P. Turban, S. Andrieu, M. Piecuch, E. Snoeck, A. Traverse, E. Foy,
  and C. Theodorescu, \epjb {\bf 15},  41  (2000).

\bibitem{bib-MUL00}
P. M{\"u}ller and R. Kern, \ss {\bf 457},  229  (2000).

\bibitem{bib-SHE97b}
J. Shen, R. Skomsky, M. Klaua, H. Jenniches, S.~S. Manoharan, and J. Jirschner,
  \prb {\bf 56},  2340  (1997).

\bibitem{bib-GAM02}
P. Gambardella, A. Dallmeyer, K. Maiti, M.~C. Malagoli, W. Eberhardt, K. Kern,
  and C. Carbone, \nature {\bf 416},  301  (2002).

\bibitem{bib-FRU06b}
F. Cheynis, N. Rougemaille, O. Fruchart, and A.~K. Schmid, \ss  (2006),
  submitted.

\bibitem{bib-CAM96}
J. Camarero, T. Graf, J.~J. {de Miguel}, R. Miranda, W. Kuch, M. Zharnikov, A.
  Dittschar, C.~M. Schneider, and J. Kirschner, \prl {\bf 76},  4428  (1996).

\bibitem{bib-BIS99}
F. Bisio, R. Moroni, M. Canepa, L. Mattera, R. Bertacco, and F. Ciccacci, \prl
  {\bf 83},  4868  (1999).

\bibitem{bib-FER00}
J. Ferr\'on, L. G\'omez, J.~M. Gallego, J. Camarero, J.~E. Prieto, V. Cros,
  A.~L. {V\'azquez de Parga}, J.~J. {de Miguel}, and R. Miranda, \ss {\bf 459},
   135  (2000).

\bibitem{bib-JUB01}
P.~O. Jubert, S. Jaren, and C. Meyer,  in {\em Magnetic storage systems beyond
  2000}, Vol.~41 of {\em NATOSCIENCE SERIES II: Mathematics, Physics and
  Chemistry}, edited by G.~C. Hadjipanayis (Kluwer Academic Publishers,
  Dordrecht, The Netherlands., 2001), experiment growth magnetism
  Self-assembled magnetic nanostructures: Fe islands on a Mo(110) surface.

\bibitem{bib-STE87}
A. Stenborg and E. Bauer, \ss {\bf 135},  394  (1987).

\bibitem{bib-STE87b}
A. Stenborg and E. Bauer, \ss {\bf 189-190},  570  (1987).

\bibitem{bib-STE89}
A. Stenborg, J.~N. Andersen, O. Bj{\"o}rneholm, A. Nilsson, and N. M{\aa}rtensson, \prl
  {\bf 63},  187  (1989).

\bibitem{bib-STE89b}
A. Stenborg, O. Bj{\"o}rneholm, A. Nilsson, N. M{\aa}rtensson, J.~N. Andersen, and C.
  Wigren, \prb {\bf 40},  5916  (1989).

\bibitem{bib-SMI96}
A.~R. Smith, K.-J. Chao, Q. Niu, and C.-K. Shih, \science {\bf 273},  226
  (1996).

\bibitem{bib-ALT97}
I.~B. Altfeder, K.~A. Matveev, and D.~M. Chen, \prl {\bf 78},  2815  (1997).

\bibitem{bib-YEH00}
V. Yeh, L. {Berbil-Bautista}, C.~Z. Wang, K.~M. Ho, and M.~C. Tringides, \prl
  {\bf 85},  5158  (2000).

\bibitem{bib-SU01}
W.~B. Su, S.~H. Chang, W.~B. Jian, C.~S. Chang, L.~J. Chen, and T.~T. Tsong,
  \prl {\bf 86},  5116  (2001).

\bibitem{bib-GAV99}
L. Gavioli, K.~R. Kimberlin, M.~C. Tringides, J.~F. Wendelken, and Z. Zhang,
  \prl {\bf 82},  129  (1999).

\bibitem{bib-HUA98}
L. Huang, S. {Jay Chey}, and J. Weaver, \ss {\bf 616},  L1101  (1998).

\bibitem{bib-LUH01}
D.~A. L.~T. Miller, J.~J. Paggel, M.~Y. Chou, and T.~C. Chiang, \science {\bf
  292},  1131  (2001).

\bibitem{bib-OTE02}
R. Otero, A.~L. {V\'azquez de Parga}, and R. Miranda, \prb {\bf 66},  115401
  (2002).

\bibitem{bib-REP02b}
V. Repain, G. Baudot, H. Ellmer, and S. Rousset, \mseb {\bf 96},  178  (2002).

\bibitem{bib-LIN04}
W.~L. Ling, T. Giessel, K. Th{\"u}rmer, R.~Q. Hwang, N.~C. Bartelt, and K.~F.
  {McCarty}, \ss {\bf 570},  L297L303  (2004).

\bibitem{bib-VOI91}
B. Voigtl{\"a}nder, G. Meyer, and N.~M. Amer, \prb {\bf 44},  10354  (1991).

\bibitem{bib-VOI91b}
B. Voigtl{\"a}nder, G. Meyer, and N.~M. Amer, \ssl {\bf 255},  L529  (1991).

\bibitem{bib-GAM03b}
P. Gambardella, \jpcm {\bf 15},  S2533  (2003).

\bibitem{bib-CHE01}
S. Cherifi, C. Boeglin, S. Stanescu, J.~P. Deville, C. Mocuta, H. Magnan, P.
  {Le F{\`e}vre}, P. Ohresser, and N.~B. Brookes, \prb {\bf 64},  184405
  (2001).

\bibitem{bib-FIG93}
J. {de la Figuera}, Prieto, C. Ocal, and R. Miranda, Phys. Rev. B {\bf 47},
  13043  (1993).

\bibitem{bib-FIG96}
J. {de la Figuera}, J.~E. Prieto, G. Kostka, S. M\"uller, C. Ocal, R. Miranda,
  and K. Heinz, \ss {\bf 349},  L139  (1996).

\bibitem{bib-PED97}
M.~O. Pedersen, I.~A. B\"onicke, E.~L. sgaard, I. Stensgaard, A. Ruban,
  J.~A.~N. rsov, and F. Besenbacher, \ss {\bf 387},  86  (1997).

\bibitem{bib-MUL05}
P. M{\"u}ller and A. Sa\'ul, \ssr {\bf 54},  157  (2005).

\bibitem{bib-MO90}
Y.~W. Mo, D.~E. Savage, B.~S. Schwartentruber, and M.~G. Lagally, \prl {\bf
  65},  1020  (1990).

\bibitem{bib-FLO99}
J.~A. Floro, E. Chason, L.~B. Freund, R.~D. Twesten, R.~Q. Hwang, and G.~A.
  Lucadamo, \prb {\bf 59},  1990  (1999).

\bibitem{bib-ROU06}
N. Rougemaille and A.~K. Schmid, \jap {\bf 99},  08S502  (2006).

\bibitem{bib-TSI03}
D.~V. Tsivlin, V.~S. Stepanyuk, W. Hergert, , and J. Kirschner, \prb {\bf 68},
  205411  (2003).

\bibitem{bib-FRU03c}
P.~O. Jubert, J.~C. Toussaint, O. Fruchart, C. Meyer, and Y. Samson, \epl {\bf
  63},  135  (2003).

\bibitem{bib-FRU04c}
O. Fruchart, J.~C. Toussaint, P.-O. Jubert, W. Wernsdorfer, R. Hertel, J.
  Kirschner, and D. Mailly, \prb {\bf 70},  172409  (2004), brief Report.

\bibitem{bib-FRU05}
R. Hertel, O. Fruchart, S. Cherifi, P.-O. Jubert, S. Heun, A. Locatelli, and J.
  Kirschner, \prb {\bf 72},  214409  (2005).

\bibitem{bib-GRA86}
U. Gradmann, J. Korecki, and G. Waller, \apa {\bf A39},  101  (1986).

\bibitem{bib-HIL87}
B. Hillebrands, P. Baumgart, and G. G\"untherodt, \prb {\bf 36},  2450  (1987).

\bibitem{bib-ELM90}
H.~J. Elmers and U. Gradmann, \apa {\bf A51},  255  (1990).

\bibitem{bib-BAU91}
P. Baumgart, B. Hillebrands, and G. Guntherodt, \jmmm {\bf 93},  225  (1991).

\bibitem{bib-ELM91}
H.~J. Elmers, T. Furubayashi, M. Albrecht, and U. Gradmann, \jap {\bf 70},
  5764  (1991).

\bibitem{bib-GRA93}
U. Gradmann,  in {\em Handbook of magnetic materials}, edited by K.~H.~J.
  Buschow (Elsevier Science Publishers B. V., North Holland, 1993), Vol.~7,
  Chap.~1, pp.\ 1--96.

\bibitem{bib-GER93}
F. Gerhardter, Y. Li, and K. Baberschke, \prb {\bf 47},  11204  (1993).

\bibitem{bib-FRI94}
H. Fritzsche, H.~J. Elmers, and U. Gradmann, \jmmm {\bf 135},  343  (1994).

\bibitem{bib-ELM94b}
H.~J. Elmers and U. Gradmann, \ss {\bf 304},  201  (1994).

\bibitem{bib-OSG95b}
R.~M. {Osgood~III}, R.~L. White, and B.~M. Clemens, \mrssp {\bf 384},  209
  (1995).

\bibitem{bib-GAL00}
I. Galanakis, M. Alouani, and H. Dreysse, \prb {\bf 62},  3923  (2000).

\bibitem{bib-QIA01}
W. Qian and W. H\"ubner, \prb {\bf 64},  092402  (2001).

\bibitem{bib-FRUunpub}
O.~F. \etal, unpublished.

\bibitem{bib-BOD04b}
M. Bode, A. Wachowiak, J. Wiebe, A. Kubetzka, M. Morgenstern, and R.
  Wiesendanger, \apl {\bf 84},  948  (2004).

\bibitem{bib-RAM98}
U. Ramsperger, A. Vaterlaus, U. Maier, and D. Pescia, \ass {\bf 130-132},  889
  (1998).

\bibitem{bib-VAN84}
H.~A.~M. {Van den Berg}, \jmmm {\bf 44},  207  (1984).

\bibitem{bib-VAN86}
H.~A.~M. {Van den Berg}, \jap {\bf 60},  1104  (1986).

\bibitem{bib-JUB01b}
P.~O. Jubert, Ph.D. thesis, Universit{\'e} Joseph Fourier, Grenoble, 2001.

\end{thebibliography}

\finalize{

\section*{Convention}

\begin{itemize}
  \item Use AL, not ML
  \item Use \textsl{preferred} height, not \textsl{constant} height.
  \item Use only Celsius degrees.
  \item Use Nanobowtie, and bow tie.
  \item use twin, not variant.
  \item Webster: utiliser seulement \textsl{unwetting} et non pas \textsl{dewetting}.
\end{itemize}

 }

\end{document}